\documentclass{article}
\usepackage{a4wide}
\usepackage{amssymb}
\usepackage{amsmath}

\newtheorem{defn}{Definition}

\newtheorem{thm}{Theorem}

\newtheorem{lem}{Lemma}

\def\sn{\mathop{\rm sn}\nolimits}
\def\cn{\mathop{\rm cn}\nolimits}
\def\dn{\mathop{\rm dn}\nolimits}

\renewcommand{\theequation}{\thesection.\arabic{equation}}
\makeatletter \@addtoreset{equation}{section} \makeatother

\title{
Functional Equations and Poincare Invariant Mechanical Systems
}

\author{J.G.B. Byatt-Smith\thanks{E-mail:Byatt@ed.ac.uk}\ \ and
H. W. Braden\thanks{E-mail:hwb@ed.ac.uk}\\
\normalsize
\em Department of Mathematics and Statistics,\\
\normalsize
\em The University of Edinburgh, \\
\normalsize
\em Edinburgh, UK \\
}

\begin{document}

\renewcommand{\thepage}{}
\begin{titlepage}

\maketitle
\vskip-9.5cm
\hskip12.4cm
MS-01-00?
\vskip8.8cm

\begin{abstract}
We study the following functional equation that has arisen in the context
of mechanical systems invariant under the Poincar\'e algebra:
\begin{equation*}
\sum\limits_{i=1}^{n+1}\dfrac{\partial }{\partial x_{i}}\prod\limits_{j\neq
i}f\left( x_{i}-x_{j}\right) =0,\qquad n \geq 2.
\end{equation*}
New techniques are developed and the general solution within a certain class
of functions is given. New solutions are found.

\end{abstract}

\begin{flushleft}
\textbf{2000 AMS Subject Classification}: Primary
39B32 
30D05  
33E05 
\end{flushleft}

\begin{flushleft}
\textbf{Key Words}: Functional Equation, Mechanics
\end{flushleft}

\vfill
\end{titlepage}
\renewcommand{\thepage}{\arabic{page}}

\section{Introduction\strut}
\renewcommand{\theequation}{\thesection.\arabic{equation}}

The differential equation
\begin{equation}
\left|
\begin{array}{ccc}
1 & 1 & 1 \\
f\left( u\right) & f\left( v\right) & f\left( w\right) \\
f^{\prime }\left( u\right) & f^{\prime }\left( v\right) & f^{\prime }\left(
w\right)
\end{array}
\right| =0,\;\text{subject to }u+v+w=0,
\label{eqn3}
\end{equation}
is one form of the first of  a series of differential equations that can be
written as
\begin{equation}
\sum\limits_{i=1}^{n+1}\dfrac{\partial }{\partial x_{i}}\prod\limits_{j\neq
i}f\left( x_{i}-x_{j}\right) =0,\qquad n \geq 2,
\label{eqgen}
\end{equation}
subject to the constraint that $f$ is an even function.
These equations, which have appeared in the context of constructing mechanical
systems with certain invariances, are the focus of this paper.
(We will later review these connections between functional equations and
mechanical systems.)
When $n=2,$ (\ref{eqgen}) gives
\begin{equation}
\dfrac{\partial }{\partial x_{1}}\left( f\left( x_{1}-x_{2}\right) f\left(
x_{1}-x_{3}\right) \right) +\dfrac{\partial }{\partial x_{2}}\left( f\left(
x_{2}-x_{3}\right) f\left( x_{2}-x_{1}\right) \right)
+\dfrac{\partial }{\partial x_{3}}\left( f\left( x_{3}-x_{1}\right) f\left(
x_{3}-x_{2}\right) \right) =0.
\label{1d3}
\end{equation}
Using the evenness of $f$ we may express this as
\begin{equation}
\dfrac{\partial }{\partial x_{1}}\left( f\left( x_{1}-x_{2}\right) f\left(
x_{3}-x_{1}\right) \right) +\dfrac{\partial }{\partial x_{2}}\left( f\left(
x_{2}-x_{3}\right) f\left( x_{1}-x_{2}\right) \right)
+\dfrac{\partial }{\partial x_{3}}\left( f\left( x_{3}-x_{1}\right) f\left(
x_{2}-x_{3}\right) \right) =0.
\label{1d4}
\end{equation}
Equation (1.1) can be written in this form if we put $u=x_{1}-x_{2},
v=x_{2}-x_{3}$ and $w=x_{3}-x_{1},$ which automatically satisfies
the constraint, although the assumption of evenness is not required.

Braden and Byatt-Smith \cite{BBS} proved, as part of a more general theorem,
that the complete solution set for the function $f$ which satisfies
(\ref{eqn3}) is
\begin{equation}
f\left( u\right) =a+bu \text{ or } a+b\;e^{cu}   \tag{1.5a}
\end{equation}
or
\begin{equation}
f\left( u\right) =a+b\,{\wp}
\left( cu+d,g_{2},g_{3}\right) .  \tag{1.5b}
\end{equation}
\setcounter{equation}{5} Here ${\wp }$ is the Weierstrass ${\wp }-$function.
Equation (1.5b) has six constants associated with it,
namely $\{ a,b,c,d,g_{2},g_{3}\}$, where $g_{2}$ and $g_{3}$ are the two
constants which relate to the two periods of ${\wp } \left( z\right)$, and
$d$ is one third of any period. However
${\wp } \left( z\right) $ satisfies the equation ${\wp }
^{\prime^{2}}=4{\wp } ^{3}-g_{2}{\wp } -g_{3},$ with $z^{2}{
\wp } \left( z\right) \rightarrow 1$ as $z\rightarrow 0.$ Hence ${
\wp } $ satisfies the scaling law ${\wp } \left(
cu,g_{2},g_{3}\right) \equiv c^{-2}{\wp } \left(
u,c^{4}g_{2},c^{6}g_{3}\right) ,$ so that without loss of generality we may
take $c=1$ in (1.5b). The solution set represented by (1.5a), which is a
subset of (1.5b), does not
require the constraint $u+v+w=0$ to be satisfied and is the general solution
of the differential equation
\begin{equation}
f^{\prime }f^{\prime \prime \prime }-f^{\prime \prime ^{2}}=0.
\label{1d6}
\end{equation}
The solution set (1.5b) only satisfies (\ref{eqn3}) provided the constraint is
satisfied and is the general solution of
\begin{equation}
f^{\prime ^{2}}+Af^{3}+Bf^{2}+Cf+D=0,
\label{1d7}
\end{equation}
or, upon eliminating the arbitrary constants, $A,B,C$ and $D$,
\begin{equation}
f^{\prime ^{2}}f^{(v)}-3f^{\prime }f^{\prime \prime }f^{(iv)}+3f^{\prime
\prime ^{2}}f^{\prime \prime \prime }-f^{\prime }f^{\prime \prime \prime
^{2}}=0.
\label{1d8}
\end{equation}
When we consider (1.5b) as the general solution of (\ref{1d7})
or (\ref{1d8}),  $d$ appears as an arbitrary  constant. However if (1.5b)  is
also to satisfy (\ref{eqn3}) then $d$ is not arbitrary  and
by substituting the solution back into (\ref{eqn3}) can be shown to be
either zero or any integer multiple of one third of any period of
 ${\wp } \left( z\right)$.
A simpler proof that all of the solutions of (\ref{eqn3}) are contained in the
solution set of (1.5), subject to the above condition on $d,$ can be
obtained by eliminating the functions $f\left( v\right) $ and $f\left(
w\right) $ by taking suitable combinations of derivatives of (\ref{eqn3}). At
various points in the proof the equation factorises to give (\ref{1d6}) or (\ref{1d7})
as factors. (See Appendix A.)

For the last 15 years the nature of the solutions to (\ref{eqgen}) has remained
open. One can show \cite{RS} that (1.5b),
with $d=0$ to ensure the evenness of $f$, satisfy equation (\ref{eqgen})
and it has been conjectured that such were the only solutions.
Though the method of eliminating functions just noted above for the case
of (\ref{eqn3}) should be applicable in the general case,
the algebra involved to completely define the
solution set is quite considerable. In the case of (\ref{eqgen}) when $n \geq 3$
the amount of algebra appears to be so large that even a Maple
calculation cannot handle the details necessary to provide a definition
of the solution set. Here we shall develop new techniques to handle the
equation. Our main result contains a surprise. To describe this let us
introduce the following:
\begin{defn}
Let $\cal{A}$ denote the class of meromorphic functions $f(z)$
whose only singularity on the real axis is a double pole at the origin,
and let ${\cal{A}}_{p}$  denote the class of periodic functions $f(z)$
whose period is $p$ (where $p$ is real) and whose only singularities on the 
real axis are double poles at $z=n\,p$, $n\in\mathbb{Z}$.
\end{defn}
\begin{thm}
For functions $f(z)\in {\cal{A}}\cup {\cal{A}}_{p}$
the  general even solution of (\ref{eqgen}) is:
\newline a) for all even $n\ge2$ given by (1.5b) with $d=0$ while\newline
b) for odd $n\ge3$ there are in addition to the solutions
(1.5b) with $d=0$ the following:
$$h_1(z)=\sqrt{(\wp(z)-e_2)(\wp(z)-e_3)}=
\frac{\sigma_2(z)\sigma_3(z)}{\sigma\sp2(z)}=
\frac{\theta_3(v)\theta_4(v)}{\theta_1\sp2(v)}
\frac{\theta_1\sp{\prime2}(0)}{4\omega^2\theta_3(0)\theta_4(0)}=
b\frac{\dn(u)}{\sn\sp2(u)},$$
$$h_2(z)=\sqrt{(\wp(z)-e_1)(\wp(z)-e_3)}=
\frac{\sigma_1(z)\sigma_3(z)}{\sigma\sp2(z)}=
\frac{\theta_2(v)\theta_4(v)}{\theta_1\sp2(v)}
\frac{\theta_1\sp{\prime2}(0)}{4\omega^2\theta_2(0)\theta_4(0)}=
b\frac{\cn(u)}{\sn\sp2(u)},$$
$$h_3(z)=\sqrt{(\wp(z)-e_1)(\wp(z)-e_2)}=
\frac{\sigma_1(z)\sigma_2(z)}{\sigma\sp2(z)}=
\frac{\theta_2(v)\theta_3(v)}{\theta_1\sp2(v)}
\frac{\theta_1\sp{\prime2}(0)}{4\omega^2\theta_2(0)\theta_3(0)}=
b\frac{\cn(u)\dn(u)}{\sn\sp2(u)}.$$
\end{thm}
Here $$\sigma_\alpha(z)=
\frac{\sigma(z+\omega_\alpha)}{\sigma(\omega_\alpha)}
e\sp{-z\zeta(\omega_\alpha)},
\qquad u=\sqrt{e_1-e_3}\,z ,
\qquad v=\frac{z}{2\omega},
\qquad b={e_1-e_3},
$$
with $\omega_1=\omega$, $\omega_2=-\omega-\omega'$ and $\omega_3=\omega'$,
and we have given representations in terms of the Weierstrass elliptic
functions, theta functions and the Jacobi elliptic functions \cite{WW}.
For appropriate ranges of $z$ the solutions are real.
These exhaust the even periodic solutions of (\ref{eqgen}) and their
degenerations yield all the even solutions with only a double pole at
$x=0$ on the real axis.
When $n=2,3$ the theorem can be proved without the assumption that
$f(z)\in {\cal{A}}\cup {\cal{A}}_{p}$ provided it is assumed that $f$ is
meromorphic with a double pole at the origin. We also conjecture this latter
assumption is all that is required for $n\ge4$ but have been unable to
prove this.
The surprise is the appearance of these new solutions for odd $n$, which in 
turn yield new Poincar\'e invariant mechanical systems.

Our paper is arranged as follows. For completeness in \S2 we will
describe the origin of (\ref{eqgen}) and of the several connections
between functional equations and mechanical systems. We then turn to
methods for finding the solution set of (\ref{eqgen}). In \S3 two methods
using a series solution introduced in Braden and Byatt-Smith \cite{BBS} are
discussed. These methods yield either a Laurent series
for the solution set or a set of differential equations, whose common
solution the solution set must satisfy. The advantage of these methods
is that they define the solution set, or the differential
equations that the set must satisfy, for (\ref{eqgen}) when $n=2$ or $n=3$.
Even though these methods appear intractable for larger values of $n$, 
the equalities derived for $n=2$ and $3$
turn out to be sufficient to define the solution set for all $n$.
In order to establish this we require a new method, valid for all $n$.
This is provided by the {\it Fourier transform method}. We develop this
method for (\ref{eqn3}) in \S4 and \S 5  and for
(\ref{eqgen}) for all $n$ in \S 6 and \S 7.
In so doing we prove theorem 1 and define completely these new set of solutions.

\section{Some Mechanical Systems}
\renewcommand{\theequation}{\thesection.\arabic{equation}}

Some years ago Ruijsenaars and Schneider \cite{RS} initiated  the study of
mechanical systems exhibiting an action of Poincar\'e algebra
\begin{equation}
\{H,B\}=P ,\qquad \{P,B\}=H,\qquad \{H,P\}=0.
\label{poincare}
\end{equation}
Here $H$ is the Hamiltonian of the system generating time-translations,
$P$ is a space-translation generator and $B$ the generator of boosts.
The models they discovered were found to posses other nice features: they were
in fact integrable and a quantum version of them naturally existed.
These models also appear in various field theoretic contexts (see \cite{BKr}).
Ruijsenaars and Schneider began with the ansatz for a system of $n+1$
particles interacting on the line,
$$
H=\sum_{j=1}\sp{n+1}\cosh p_j\, \prod_{k\ne j}F(x_j-x_k)
,\qquad
P=\sum_{j=1}\sp{n+1}\sinh p_j\, \prod_{k\ne j}F(x_j-x_k)
$$
and
$$B=-\sum_{j=1}\sp{n+1}x_j .  $$
With this ansatz and the canonical Poisson bracket $\{x_i,p_j\}=\delta_{ij}$
the first two Poisson brackets of (\ref{poincare}) involving the boost
operator $B$ are automatically satisfied.
Supposing further that $F(x)=F(-x)$, then the final Poisson bracket is
equivalent to the functional equation
\begin{equation}
\{H,P\}=0 \Longleftrightarrow
\sum_{j=1}\sp{n+1}\partial_j \prod_{k\ne j}F\sp2(x_j-x_k)=0,
\label{functional}
\end{equation}
With $f(x)=F\sp2(x)$ this is precisely (\ref{eqgen}), and so solutions of this
equation (with even square root) yield Poincar\'e invariant mechanical systems. 
At the time
Ruijsenaars and Schneider were able to to show (1.5b) (with $d=0$) gave
solutions to these equations for all $n$.
These solutions in fact yield $n+1$ independent, mutually Poisson commuting
conserved quantities, and so are a completely integrable mechanical system.
A scaling limit of the Ruijsenaars-Schneider model
yields the Calogero-Moser system with Hamiltonian
\begin{equation}
H= \frac{1}{2} \sum\limits_{i=1}^{n+1}p_{i}^{2}+
\frac{1}{6}\sum\limits_{i\neq j}\wp\left( q_{i}-q_{j}\right)
\end{equation}
which is another well-studied completely integrable system \cite{vDV}.

We note that many connections exist between functional equations
and integrable quantum and classical systems \cite{BB1, BB2, BCb, bp, BK,
Ca2, DFS, Gu}.
Functional equation (\ref{eqn3}) (without any assumptions on the parity of the
function $f$) arises for example when characterising quantum mechanical
potentials whose ground state wavefunction (of a given form) is factorisable
\cite{cal, suth}.
More recently it has been shown to characterise the Calogero-Moser system
\cite{Bra}.
Several functional equations appearing in this setting and whose general
solutions have still to be found are given in \cite{BB2}.

\newpage
\noindent
\section{Series Solution Approaches}
\renewcommand{\theequation}{\thesection.\arabic{equation}}
We shall now describe two methods based on a series approximation
for studying (\ref{eqgen}).

{\bf Method 1:} Obtaining a Series Solution.

One method of attempting to prove the conjecture is to assume all the
$x_{i}, i=1\cdots n+1,$ are small and of the same order, so that we write
write $x_{i}=t\zeta_i$.
Then we assume that all the even functions, such as  $f\left( t\zeta \right) $
can be expressed as a power series in $t$ with $\zeta$ as an order one
parameter, in the form
\begin{equation}
f\left( t\zeta \right) =\sum_{j=0}^{\infty }a_{j}c_{2j-2}\left( t\zeta \right)
^{2j-2}.
\label{3d1}
\end{equation}
The constants $c_{j}$ are given by  $c_{j}=1$, if $j <0$ and $c_{j}=1/j!,$
if $j \geq 0$ and are included for convenience.
The series (\ref{3d1}) allows for a double pole at the origin which can easily
be shown to be the only allowable singularity. The coefficients $a_{j}$ are
determined by equating to zero the coefficients of the powers of $t^{2j}$ in the
subsequent expansion of (\ref{eqgen}). These coefficients are of course
functions of $\zeta _{i}$ as well as $a_{j}$. However each coefficient
factorises into a
product of homogeneous polynomials in $\zeta _{i},$ independent of $a_{j},$
and a factor dependent on the $a_{j}$ only. Equating this coefficient to
zero successively determines $a_{j}$ for all $j\geq 4$ in terms of $
a_{0},a_{1},a_{2},$ and $a_{3}$ which are arbitrary. This process can be
completed to any desired order, $J,$ if the expansion (\ref{3d1}) is truncated
at a suitable finite value. Substitution of this finite polynomial into
(\ref{1d7}) shows for the cases where the method works, that for a suitable
choice of $ \left\{ A,\;B,\;C,\;D\right\} $, (\ref{1d7}) can be satisfied to
any desired order.

The proof of the above statement is, of course, incomplete: the form of
the general term, $a_{j},$ is not obtained and hence we cannot show that the
full expansion (\ref{3d1}) satisfies (\ref{1d7}). This method works for $n=2$
and $n=4,$ but for values of $n\geq 5$ the amount of algebra involved becomes
so large that even a Maple calculation cannot handle the details. When $n=3$,
this method does not work completely in that it leaves $a_{4}$ arbitrary.
Subsequent methods show that this is not the case and we will resolve this
difficulty later.

It is interesting to note that the same procedure works for solutions of
(\ref{eqn3}), although the assumption of evenness is not required. Hence we
look for a solution
\begin{equation}
f\left( t\zeta \right) =\sum_{j=0}^{\infty }a_{j}c_{j-2}\left( t\zeta \right)
^{j-2}.
\label{3d2}
\end{equation}
If we take $a_{0}\neq 0$ the process automatically gives $a_{2j+1}\equiv 0$
producing the same even function as obtained via (\ref{3d1}) for the solutions
of (\ref{eqgen}). However, if we take $a_{0}=0$ then we find that
$a_{1}\equiv 0$ and (\ref{3d2}) is then a Taylor series.
We proceed as above to produce the coefficients $a_{j}$ for all $j\geq 7$ in
terms of $a_{2}$ to $a_{6}$ and show that (\ref{1d7}) can be satisfied to any
order.  However, $a_{5}$ is not arbitrary and is given by
\begin{equation}
a_{3}\;a_{5}=a_{4}^{2}.
\label{3d3}
\end{equation}
This condition is automatically satisfied for all functions in the set (1.5a).
However (\ref{3d3}) is also equivalent to the condition
${\wp } ^{\prime \prime ^{2}}\left( d\right) =12{\wp }
^{\prime ^{2}}\left( d\right) /{\wp } \left( d\right), \; $
which is satisfied when $d$ is one third or two thirds of any period
of ${\wp } \left( z\right) .$ This gives the required condition on
the constant $d$ appearing in (1.5b), when $f$ belongs to this
solution set.

{\bf Method 2:} Obtaining a Series of Differential Equations.

An alternative method is to assume that one variable, for example
$x_{1}$ is not small and write $x_{1}=x$ and $x_{i}=t\zeta _{i},$ $2\leq i\leq
n+1 $ together with (\ref{3d1}), although $x_{1}$ can be replaced by any linear
combination of the $x_{i},$ so long as $n$ variables are scaled by $t.$
The expansion of $f(x-t\zeta )$ then naturally produces coefficients of
$t^{j},$ which are function of $f$ and its higher derivatives.
The coefficients of $t^{j},$ in the expansion of (\ref{eqgen}) when equated
to zero, now yield differential
equations which must be satisfied by $f\left( x\right) $ but contain the
``arbitrary'' constants, $a_{j}.$

For the case $n=2$ the first equation is
\begin{equation}
a_{0}f^{\prime \prime \prime }+12a_{1}f^{\prime }-12ff^{\prime }=0.
\label{3d4}
\end{equation}
This proves to be sufficient to determine a differential equation for $f$ in
the sense that the elimination of the constants $a_{0}$ and $a_{1}$ by
differentiation gives (\ref{1d8}). The higher order coefficients of $t^{j}$
produce equations similar to (\ref{3d4}) but contain the constants $a_{j}$ with
$ j\geq 4$ which are not arbitrary. In this case eliminating all the constants
by differentiation will yield an equation which is still necessary but not
sufficient to determine $f$. If we seek a series solution to (\ref{3d4}) by
looking for a series solution of the form $f(x)=\sum_{j=0}^{\infty
}b_{j}c_{2j-2}x^{2j-2}$ then we obtain $b_{0}=a_{0}$ and $ b_{1}=a_{1}$,
with $b_{2}$ and $b_{3}$ arbitrary. The coefficients $b_{j},j \geq 4$
are then determined by these four constants, the recurrence relation
being the same as in method 1. The equations which contain $a_{2}$ and
$a_{3}$ also require $b_{2}=a_{2}$ and $ b_{3}=a_{3}$ and so also
reproduce the series (\ref{3d1}).

This method is more complete than the previous method in that it does yield
a necessary differential equation, namely (\ref{1d8}), that all even
solutions of (\ref{eqgen}) for $n=2$ must satisfy. However we cannot prove that
all the differential equations produced by the expansion are satisfied.
However this is not a problem since it is easy to verify, by substitution, that
all even solutions of (1.5b) satisfy (\ref{eqgen}) for this case.

Again the method can be adapted to obtain the solution of (\ref{eqn3}) which are
not even, using (\ref{3d2}) with $a_{0}=a_{1}=0$. The procedure is the same
except that because $f\left( \zeta \right) $ is not even we obtain differential
equations containing $f\left( \zeta \right) $ and $f\left( -\zeta \right) .$
When one of these functions is eliminated we can then show that either
(\ref{1d6}) or (\ref{1d8}) is satisfied. However, the differentiations required
to eliminate the constants $a_{j}$ means that the necessary condition
(\ref{3d3}) is not recovered.

For the case $n=3$ the situation is more complicated since the series of
equations similar to (\ref{3d4}) only produce necessary equations for $f.$ In
order to show that $f$ satisfies (\ref{1d7}) we need to take two equations
similar to (\ref{3d4}) and find the set of solutions common to both differential
equations. The algebraic details are quite complicated and require a Maple
calculation. The details are not repeated here but a summary of the results
is given. A copy of the Maple programme which produces these results with
further explanation can be obtained by contacting the first author.

The two equations, whose common solution satisfies (\ref{eqgen}), for $n=3$ are
given by (A1.1) and (A1.2) in appendix A. To determine the equation or
equations for these common solutions we effectively eliminate the arbitrary
constants $a_{0},a_{1},a_{2}$ and $a_{3}.$ These constants are the ones that
appear in the expansion (\ref{3d1}) used to derive (A1.1) and (A1.2).

Before we derive these equations for the common solution we investigate the
form of these solutions by looking for a common solution of the form (\ref{3d1})
but with the constants $a_{j}$ replaced by $b_{j}$. The result again
shows that $b_{0}=a_{0}, b_{1}=a_{1},b_{2}=a_{2}$ and $ b_{3}=a_{3}$.
However, initially $b_{4}$ is not determined but $b_{5}$ and $b_{6}$ are
determined in terms of $b_{j}, j \leq 4$. However we get two different
values of $b_{7}$
and equating these values give an equation which has two solutions, namely
\begin{equation}
b_{3}=\dfrac{60\left( 2b_{1}^{2}-b_{0}b_{2}\right) }{7b_{0}^{2}}
\mbox{ or }
b_{4}=\dfrac{60b_{2}^{2}}{b_{0}}\;. \tag{3.5a,b}
\end{equation}
\setcounter{equation}{5}If we choose (3.5b) then in turn we find both equations require a common
value for $b_{j}$ for $j\geq 7$ and the series generates, as in the previous
method, a solution of (\ref{1d7}) or (\ref{1d8}) and so belongs to the solution set (1.5).

However, if we choose (3.5a) to be satisfied then the two equations have
different solutions for $b_{8}$ and equating these values determines $b_{4}.$
This value differs from that given in (3.5b) and hence generates a
solution $f\left( z\right) =f_{1}\left( z\right) $, with three arbitrary
constants, which is not a solution of (\ref{1d7}) or (\ref{1d8}), unless
\begin{equation}
b_{2}=\dfrac{6b_{1}^{2}}{5b_{0}}\;.
\label{3d6}
\end{equation}
It is easy to verify that when (\ref{3d6}) is satisfied,  any finite
truncation of the  series is equal to the same truncation of the solution
\begin{equation}
f_{2}\left( z\right) =b_{1}{\wp } \left( \sqrt{\dfrac{b_{1}}{b_{0}}}
z,12,8\right) +b_{1}.
\label{3d7}
\end{equation}
This is a subset of the general solution of (\ref{1d8}), which may be
expressed as
$$ f_{2}\left( z\right) =3b_{1}/\sin ^{2} \left( \sqrt{\dfrac{3b_{1}}{b_{0}}}
z\right) ,\qquad{\text or}\qquad
f_{2}\left(
z\right) =-3b_{1}/\sinh ^{2} \left( \sqrt{-\dfrac{3b_{1}}{b_{0}}}z \right),
$$
depending on whether $\dfrac{3b_{1}}{b_{0}}$ is greater than, or less than, zero
respectively.
Since the choice of (3.5a) requires a specific choice of $b_{3}$ and hence $
a_{3}$ it may be thought that this solution fails to be a solution of the full
equation (\ref{eqgen}). However, as we will prove, this supposition proves to be
erroneous.

We can also determine the equation for the common solution of (A1.1) and
(A1.2) by effectively eliminating the arbitrary constants $a_{0},a_{1},a_{2}$
and $a_{3}.$ The details are contained in the Maple programme referred to
earlier and are not given here. This results in equation (A1.3) which
factorises, so that $f$ must satisfy one of the following equations
\begin{equation}
f^{\prime }=0\;\;\text{or\ }f=\text{constant,}
\label{3d8}
\end{equation}
clearly a solution, but not one of interest, and
\begin{equation}
f^{\prime \prime \prime }f-f^{\prime }f^{\prime \prime }=0\;\;\text{or\ }
f^{\prime \prime }/f=\text{constant.}
\label{3d9}
\end{equation}
We find that this equation only satisfies (A1.1) and (A1.2) if the constant
appearing in (\ref{3d9}) is zero. Thus the only even solution again is $f=$
constant. The third factor we recognise as equation (\ref{1d8}) with solution
(1.5b), with $d=0$ for an even solution. The last factor is seen to be a
fifth order ordinary differential equation which is cubic in $f^{(v)}.$ It
is easily verified that the solution $f=f_{1}\left( z\right) $ defined
earlier by taking the choice of (3.5a) for $b_{3}$ satisfies this equation,
(A1.4).

At first sight it would seem unlikely that we could obtain the most general
solution of (A1.4), which is even and has a double pole at $z=0.$ However, in
\S7 we show that the Fourier transform of the non-periodic solution is
simple and easily inverted to give
\begin{equation}
f\left( z\right) =a\cosh bz/\sinh^{2}bz\;,
\label{3d10}
\end{equation}
for arbitrary constants $a$ and $b.$ This coincides with the series for $
f_{1}\left( z\right) $ obtained earlier as the solution of (A1.4) by
choosing $a=6b_{1}$ and $b=\sqrt{{6b_{1}}/{b_{0}}},$ although this
requires the choice $b_{2}=-{7ab^{2}}/{60}\equiv -{21b_{1}^{2}}/{
5b_{0}}.$ With this knowledge, one may guess that the periodic
solution are of the form $a\;\mbox{cn}\left( bz,k\right) /\mbox{sn}
^{2}\left( bz,k\right)$ $\text{\ }a\;\mbox{dn}\left( bz,k\right) /\mbox{sn}
^{2}\left( bz,k\right) \text{ or } a\;\mbox{cn}\left( bz,k\right) \mbox{dn}
\left( bz,k\right) /\mbox{sn}^{2}\left( bz,k\right)$, or the alternate
forms given in theorem 1. These are all periodic
equivalents of (\ref{3d10}), where cn, dn and sn are the Jacobian elliptic
functions and k the modulus. It is then possible to verify, by
substituting into (A1.4)  that the most
general solution of this equation of the form required may be expressed as
\begin{equation}
f_{1}\left( z\right) =\sqrt{b_{0}^{2}W^{2}\left( z\right)
+2b_{0}b_{1}W\left( z\right) -b_{1}^{2}+\dfrac{5}{3}b_{0}b_{2},}
\label{deff}
\end{equation}
where
\begin{equation}
W\left( z\right) ={\wp } \left( z,-\dfrac{20\left(
b_{0}b_{2}-3b_{1}^{2}\right) }{3b_{0}^{2}},\dfrac{8 b_{1}\left(
5b_{0}b_{2}-3b_{1}^{2}\right) }{3 b_{0}^{3}}\right) .
\label{3d12}
\end{equation}
The quadratic in (\ref{deff}) has a double root when $6b_{1}^{2}=5b_{0}b_{2}$
which corresponds to the condition (\ref{3d6}) which gives the function $
f_{2}\left( z\right) ,$ defined by (\ref{3d7}). Thus $f_{2}\left( z\right) $ is
the two parameter family of common solution to (\ref{1d8}) and (A1.4), which
includes the solution $b_{0}z^{-2}$ in the limit $b_{1}\rightarrow 0.$

If the equation for the ${\wp } $ function is written as
\begin{equation}
{\wp } ^{\prime ^{2}}=4{\wp } ^{3}-g_{2}{\wp }
-g_{3}\equiv 4\left( {\wp } -e_{1}\right) \left( {\wp }
-e_{2}\right) \left( {\wp } -e_{3}\right) ,
\label{3d13}
\end{equation}
in the usual notation, then the quadratic in (\ref{deff}) divides the cubic on
the right hand side of (\ref{3d13}). When $e_{1}$ and $e_{3}$ are the common
roots (\ref{deff}) can be simplified to give the solution
$f_{1}\left( z\right) =a\
\mbox{cn}\left( bz,k\right) /\mbox{sn}^{2}\left( bz,k\right) $ for suitable
choices of $a,b$ and $k$ in terms of $b_{0},b_{1}$ and $b_{2}.$ When $e_{2}$
and $e_{3}$ are the common roots we recover the solution $f_{1}\left(
z\right) =a\ \mbox{dn}\left( bz,k\right) /\mbox{sn}^{2}\left( bz,k\right) $
while the third choice $\left( e_{1},e_{2}\right) $ gives $f_{1}=a\ \mbox{cn}
\left( bz,k\right) \mbox{dn}\left( bz,k\right) /\mbox{sn}^{2}\left(
bz,k\right)$.


\section{The Fourier Transform Method}

We now look at the problem of determining the solution of (\ref{eqn3}) and (
\ref{eqgen}) using Fourier transforms. There are a variety of difficulties
which are not immediately apparent. These will be treated as they arise. They
include requiring generalised Fourier transforms for functions of $x$ which are
unbounded either at infinity or at points on the real axis, and the
consideration of distributional solutions both of (\ref{1d3}), (\ref{1d4})
and of the transformed equation.

We first assume that the function $f$ in (\ref{1d4}) is not necessarily even
but is bounded on the real axis.  Then we define the double Fourier Transform
$\widehat F(k,l)\equiv T[ F ] (k,l)$ of $F(x_{1},x_{2})$ by
\begin{equation}
\widehat F(k,l) =\iint\limits_{\mathbb{R}\sp{2}}F\left(
x_{1},x_{2}\right) e^{-ikx_{1}-ilx_{2}}dx_{1}dx_{2}.
\label{4d1}
\end{equation}
With
\begin{multline}
F\left( x_{1},x_{2}\right) =\dfrac{\partial }{\partial x_{1}}\left( f\left(
x_{1}-x_{2}\right) f\left( x_{3}-x_{1}\right) \right) +\dfrac{\partial }{
\partial x_{2}}\left( f\left( x_{2}-x_{3}\right) f\left( x_{1}-x_{2}\right)
\right) \\
+\dfrac{\partial }{\partial x_{3}}\left( f\left( x_{3}-x_{1}\right) f\left(
x_{2}-x_{3}\right) \right) ,
\end{multline}
we have
\begin{multline}
\widehat F(k,l) =ike^{-i(k+l)x_{3}}\widehat{f}\left( -l\right) \widehat{f}\left(
-l-k\right) +ile^{-i\left( k+l\right) x_{3}}\widehat{f}\left( -k\right) \widehat{f}
\left( k+l\right)   \\
-i\left( k+l\right) e^{-i\left( k+l\right) x_{3}}\widehat{f}\left(- k\right) \widehat{
f}\left( l\right) ,
\end{multline}
where $\ \widehat{f}\left( k\right) =\int_{-\infty }^{\infty }f\left(
x_{1}\right) e^{-ikx_{1}}dx_{1}$ is the Fourier transform of $f\left(
x_{1}\right) .$
When $f$ and $f^{\prime }$ are bounded on the real axis it is clear that all
terms in $F\left( x_{1},x_{2}\right)$ are also bounded for all
$(x_1,x_2)$ and hence $F\left( x_{1},x_{2}\right)\equiv 0,$ for all such
bounded solutions
$f$. Thus the appropriate equation to determine $\widehat{f}\left(
k\right)$ is $\widehat F(k,l) =0$, which  gives
\begin{equation}
k\widehat{f}\left( -l\right) \widehat{f}\left( -k-l\right) +l\widehat{f}\left( k\right)
\widehat{f}\left( k+l\right) -\left( k+l\right) \widehat{f}\left( -k\right) \widehat{f}
\left( l\right) =0.
\label{4d4}
\end{equation}

However when $f$ has singularities on the real axis $F\left(
x_{1},x_{2}\right)$ is not defined at all points in the $( x_{1},x_{2})$
plane and we find that $F\left( x_{1},x_{2}\right)=0$, everywhere except
on a set of measure zero.  In particular for the even solutions of
(1.5b) this set of points is the lattice $x_{1}=x_{3} \mod(\lambda )$ and
$x_{2}=x_{3} \mod(\lambda )$, where $\lambda $ is the real period of the
function $f$.  For such functions $f$ we show, in \S 5,  that  $F\left(
x_{1},x_{2}\right) \not\equiv 0$.  We also show in \S 5 that this means that
$\widehat F =0$ is not the appropriate equation and derive the correct
equation.

Equation (\ref{4d4}) is a rather complicated functional equation when
$\widehat{f }\left( -k\right) \neq \widehat{f}\left( k\right)$.
A Taylor series method produces a three parameter family of solutions. Two
parameters are as a consequence of the fact that if \^{g}$\left( k\right) $
is a solution so is
$\widehat{f}\left( k\right) =$ $a\widehat{g}\left( bk\right) $
for all constants $a$ and $b$. This is as a result of the scaling symmetries
of the original equation (\ref{4d4}). So essentially there is a one parameter
family of solutions. However, it is not easy to recognise the solution from
its series. The method which appears to give the most simple solution is the
following. We decompose $\widehat{f}$ into an even and odd function of the
form
\begin{equation}
\widehat{f}\left( k\right) =f_{1}\left( k\right) +tkf_{2}\left( k\right) ,
\end{equation}
where $f_{1}$ and $f_{2}$ are even functions of $k$ and $t$ can be $\pm 1.$
Substituting this expression into (\ref{1d4}) yields an equation of the form
\begin{equation}
A\left( k,l\right) +tB\left( k,l\right) +t^{2}C\left( k,l\right) =0
\label{4d6}
\end{equation}
and since $t$ can be either $\pm 1$ this gives two equations
\begin{equation}
A+C=0\;\;\text{and\ }B=0.
\label{4d7}
\end{equation}

Now we assume that $l$ is small and expand (\ref{4d7}a,b) as a power series in
$l$.  Equating the coefficients of the powers of $l$ to zero gives a series of
differential equations involving $f_{1}\left( k\right) $ and $f_{2}\left(
k\right)$.
>From the first two equations we obtain
\begin{equation}
2\left( f_{2}\left( k\right) -a_{2} \right)f_{1}\left( k\right)
=k\;a_{1}\;f_{2}^{\prime }\left( k\right)
\label{4d8}
\end{equation}
and
\begin{equation}
ka_{1}f_{1}^{\prime }\left( k\right) +f_{1}^{2}\left( k\right)
+k^{2}f_{2}^{2}\left( k\right) -f_{1}\left( k\right)
a_{1}+2k^{2}a_{2}f_{2}\left( k\right) =0,
\label{4d9}
\end{equation}
where
\begin{equation}
a_{1}=f_{1}\left( 0\right) \mbox{ and }a_{2}=f_{2}\left( 0\right) .
\end{equation}
Eliminating $f_{1}\left( k\right) $ we obtain
\begin{equation}
a_{1}^{2}\left( 2\left( f_{2}-a_{2}\right) f_{2}^{\prime \prime }
-f_{2}^{\prime ^{2}}\right) +4f_{2}\left(
f_{2}^{3}-3a_{2}^{2}f_{2}-2a_{2}^{3}\right) =0
\end{equation}
and one integration yields
\begin{equation}
3a_{1}^{2}f_{2}^{\prime ^{2}}=4\left( a_{2}-f_{2}\right) \left(
ca_{2}^{3}+f_{2}^{3}+3a_{2}f_{2}^{2}\right) ,
\label{4d12}
\end{equation}
where $c$ is an arbitrary constant. This constant is the third parameter
referred to above. Examination of the cubic in (\ref{4d12}) shows that for all
(real) constants $c$ other than $c=0$ or $-4$ we have an oscillatory
solution for $f_{2}.$ These solutions must be rejected on the grounds that
if the original function $f\left( x\right) $ is bounded, $\widehat{f}\left(
k\right) $ must tend to zero as $k\rightarrow \pm \;\infty .$ When $c=-4,$ $
f_{2}\equiv a_{2}$ and (\ref{4d8}) is automatically satisfied and
(\ref{4d9}) then yields
\begin{equation}
f_{1}(k)=\sqrt{3}a_{2}k\cot\left( \sqrt{3}a_{2}k/a_{1}\right) .
\label{4d13}
\end{equation}

Again this does not represent the Fourier transform of a bounded function $
f\left( x\right) .$ However, it does illustrate the general feature of all
periodic solutions of (\ref{4d12}). When $c\neq 0$ the solution of
(\ref{4d12}) with $ f_{2}\left( 0\right) =a_{2}$ is an
elliptic function which is even. Clearly,
when $c\neq -4, \;f_{2}(k)-a_{2} $ has zeros when $k=0$ or an integer
multiple of the period of $f_{2}.$ Since (\ref{4d12}) implies that $
f_{2}^{\prime }=0$ and $f_{2}^{\prime \prime }\neq 0$ when $f_{2}=a_{2},$ at
each zero other than $k=0$, $f_{1}(k)$ has a simple pole. Hence $f_{1}$ has
a periodic array of poles, apart from $k=0$ where the singularity is
removable. This feature is illustrated by the function appearing in
(\ref{4d13}).

When $c=0$ we can solve (\ref{4d12}) with $f_{2}\left( 0\right) =a_{2}$ to get
\begin{equation}
f_{2}(k)=\dfrac{3a_{2}}{2\cosh 2\alpha k+1},
\end{equation}
where $\alpha =a_{2}/a_{1},$ so that
\begin{equation}
f_{1}(k)=\dfrac{3a_{2}k\cosh \alpha k}{\sinh \alpha k\left( 2\cosh \alpha
k+1\right) }.
\end{equation}

The two possibilities for $\widehat{f}\left( k\right) $ are then
\begin{equation}
\widehat{f}\left( k\right) =f_{1}\left( k\right) \pm kf_{2}\left( k\right) ,
\label{4d16}
\end{equation}
giving
\begin{equation}
\widehat{f}_{1}\left( k\right) =\dfrac{6a_{2}ke^{4\alpha k}}{e^{6\alpha k}-1}\;\;
\text{and }\widehat{f}_{2}\left( k\right) =\dfrac{6a_{2}ke^{2\alpha k}}{
e^{6\alpha k}-1}\;.
\end{equation}
If we normalise by choosing $6a_{2}=\pi a_{1}$ and $a_{1}=-2,$ then $\widehat{f}
_{1}$ is the Fourier transform of the function
$1/\sinh^{2}\left( x-i\pi /3\right) $ and $\widehat{f}_{2}$ is the Fourier
transform of $1/\sinh^{2}\left( x+i\pi /3\right) .$ Apart from the addition of
a constant these are ${\wp } $
functions, with periods $\left( \infty ,\pi \right) ,$ and a double pole at
the origin.

Thus, apart from the double pole at the origin and the
shift by one third and two thirds of the
imaginary period of the ${\wp } $ function there is a unique solution of
(\ref{1d4}) which is bounded on $\left( -\infty ,\infty \right) $ and tends to
zero at $\infty .$
An interesting limit is $a_{2}\rightarrow 0$ which gives $f_{2}\equiv 0$ and
$f_{1}=a_{1}.$ This is also seen to be the only solution of (\ref{4d8}) and
(\ref{4d9}) when $a_{2}=0,$ subject to the condition that $f_{1}\left( 0\right)
=a_{1}\;\;f_{2}\left( 0\right) =a_{2}=0$ and $f_{2}\rightarrow 0$ as $
k=\infty$. The inverse of the Fourier transform $\widehat{f}\left( k\right)
\equiv a_{1}$ is  no longer a function, but the distribution $a_{1}\delta \left(
x\right) $ which can be viewed as the limit
$$a_{1}\delta \left(x\right)=\lim_{\alpha \rightarrow 0,}
-\frac{a_{2}}{\alpha
\sinh^{2}\left( -x/\alpha +\pi i/3\right) }, $$
where $\pi a_{1}\alpha =6a_{2}.$

The only other regular solutions of (\ref{1d4}), on the real axis, are ones that
are either not bounded at $\infty $ or are oscillatory. Such functions have
distributional Fourier transforms. For example, the Fourier transform of $
e^{i\alpha x}$ is $2\pi \delta \left( k-\alpha \right) $ and as a
distribution $f=2\pi \delta \left( k-\alpha \right) $ satisfies (\ref{4d4}),
since
\begin{multline}
4\pi ^{2}k\delta \left( -l-\alpha \right) \;\delta \left( -k-l-\alpha
\right) +4\pi ^{2}l\delta \left( k-\alpha \right) \delta \left( k+l-\alpha
\right) \\
-4\pi ^{2}\left( k+l\right) \delta \left( -k-\alpha \right) \delta \left(
l-\alpha \right) \equiv 0.
\end{multline}
This is because $k\;\delta \left( -l-\alpha \right) \;\delta \left(
-k-l-\alpha \right) $ is zero unless $l+\alpha =l+k+\alpha =0$ or $
k=0,\;l=-\alpha ,$ when the coefficient of the product of $\delta $
functions is zero. A more formal proof can be obtained by defining the inner
products $\langle u,v\rangle _{k,l}$ and $\langle u,v\rangle _{k}$ by
\begin{equation}
\langle u,v\rangle _{k,l}\equiv \iint\limits_{\mathbb{R}^{2}}u\;v\;dkdl
\qquad \mbox{and}\qquad
\langle u,v\rangle _{k}\equiv \int\limits_{\mathbb{R}}u\;v\;dk,
\label{4d19}
\end{equation}
with a similar definition for $\langle
u,v\rangle _{l}.$
Then
\begin{eqnarray}
\langle k\delta \left( -l-\alpha \right) \delta \left( -k-l-\alpha \right)
,\;F\left( k,l\right) \rangle _{k,l}
&=&\langle \delta \left( -l-\alpha \right) \delta \left( -k-l-\alpha \right)
,\;\;kF\left( k,l\right) \rangle _{k,l}  \notag \\
&=&\langle \delta \left( -l-\alpha \right) ,\;\;-\left( l+\alpha \right) \;F\left(
-l-\alpha ,l\right) \rangle _{l} \\
&=&0.  \notag
\end{eqnarray}
Although this only shows that $e^{i\alpha x}$ is a solution of (\ref{1d4}) by
analogy we can also have $e^{\alpha x}$ although this does not have a
Fourier transform, except formally by analytic continuation.

The Fourier transform of $x$ is $2\pi i\delta ^{\prime }(k)$ and using
(\ref{4d19}) it is straightforward to prove that $k\delta ^{\prime }\left(
-l\right) \delta ^{\prime }\left( -k-l\right) +l\delta ^{\prime }\left(
k\right) \delta ^{\prime }\left( k+l\right) $ and $\left( k+l\right) \delta
^{\prime }\left( -k\right) \delta ^{\prime }\left( l\right) $ are identical
distributions. For example
\begin{eqnarray}
\langle \left( k+l\right) \delta ^{\prime }\left( -k\right)\delta ^{\prime }\left(
l\right) ,\;\;F\left( k,l\right) \rangle _{k,l} & = & \langle \delta ^{\prime } \left( -k\right)
\delta ^{\prime }\left( l\right) ,\left( k+l\right) F\left( k,l\right)
\rangle _{k,l}  \notag \\
& = & \langle \delta ^{\prime }\left( -k\right), -F\left( k,0\right) -kF_{l}\left(
k,0\right) \rangle _{k} \\
&= & -f_{k}\left( 0,0\right) -f_{l}\left( 0,0\right) .\notag
\end{eqnarray}
A similar calculation shows that
\begin{equation}
\langle k\delta ^{\prime }\left( -l\right) \delta ^{\prime }\left( -k-l\right)
,\;F\left( k,l\right) \rangle _{k,l} +\langle l\delta ^{\prime }\left( k\right) \delta
^{\prime }\left( k+l\right) ,\;F\left( k,l\right) \rangle _{k,l}\;
=-f_{k}\left( 0,0\right) -f\left( 0,0\right) .
\end{equation}

In addition we can also see that if $\widehat{f}\left( k\right) $ is not
continuous $\widehat{f}\left( 0\right) $ can be arbitrary since (\ref{4d4}) is
satisfied automatically when either $k,\;l$ or $k+l$ equals zero. In
particular, we can add an arbitrary multiple of $\delta (k)$ to any solution
of (\ref{4d4}). This corresponds to adding a constant to any solution of
(\ref{1d4}).

The above distributional solutions cover the solutions of (1.5a) that do not
satisfy (1.5b). To cater for the non even periodic solution of (1.5b) that
are also bounded on the real axis, we assume that the solutions $f\left(
x\right) $ are $2\pi $ periodic and write
\begin{equation}
f\left( x\right) =\dfrac{1}{2\pi}\sum\limits_{p=-\infty }^{\infty }
a_{p}e^{ipx}.
\end{equation}
The factor $2\pi $ is introduced so that the Fourier Transform takes the form
\begin{equation}
\widehat{f}\left( k\right) =\sum\limits_{p=-\infty }^{\infty } a_{p}\delta
\left( k-p\right) .
\label{4d24}
\end{equation}
Introducing this expression into (\ref{4d4}) and recognising that
$\widehat{f}(k)$ is only non zero when $k$ is an integer, we obtain
\begin{eqnarray}
ka_{-L}\delta \left( -l+L\right) a_{-K-L}\delta \left( -k-l+K+L\right)
&+& l\, a_{K}\delta \left( k-K\right) a_{K+L}\delta \left( k+l-K-L\right)
 \notag \\
-\left( k+l\right) a_{-K}\delta \left( -k+K\right) a_{L}\delta \left(
l-L\right) &=&0,
\label{4d25}
\end{eqnarray}
at all integers values of $k$ and $l.$
This is a distribution supported at $k=K$ and $l=L$ and is identically zero
if
\begin{equation}
Ka_{-L}a_{-K-L}+La_{K}a_{K+L}-\left( K+L\right) a_{-K}a_{L}=0,
\label{4d26}
\end{equation}
for all integer values of $\left( K,L\right) .$ Since $\widehat{f}_{1}(k)$ and
$\widehat{f}_{2}\left( k\right) $ defined by (\ref{4d16}) satisfy the continuous
version of (\ref{4d26}), it is clear that the solution
$a_{K}=\widehat{f}_{1}\left(
K\right) ,$ $K\neq 0$ with $a_{0}$ arbitrary, will satisfy (\ref{4d26}). It is
also easy to construct the corresponding function $f\left( x\right) $ that
satisfies (\ref{1d4}). If $f_{1}\left( x\right) $ is the function whose Fourier
transform is $\widehat{f}_{1}\left( k\right) $ then we define
\begin{equation}
g\left( x\right) =\sum\limits_{p=-\infty }^{\infty }f_{1}
\left( x-2\pi p\right) .
\end{equation}

This function is $2\pi $-periodic and has Fourier series $
\dfrac{1}{2\pi } \sum\limits_{K=-\infty }^{\infty }a_{K}e^{iKx}$ where
\begin{equation}
a_{K}=\int_{0}^{2\pi }\sum\limits_{p=-\infty }^{\infty
}f_1\left( x-2p\pi \right) e^{-iKx}dx
=\int_{-\infty }^{\infty }f\left( x\right) e^{-iKx}dx
=\widehat{f}_{1}\left( K\right).
\label{4d28}
\end{equation}
The above solution for $a_{K}$ can also be obtained directly from
(\ref{4d26}) by
successively solving all equations with $1\leq \left| K\right| \leq N, \;
1\leq \left| L \right| \leq N$ for $N=1,2,3,...$, the equations where $K=0
$ or $L=0$ being automatically satisfied. Apart from the fact that $a_{0}$
is undetermined and is thus arbitrary, this process, as in the continuous
case, produces a three parameter family of solutions with $a_{-2},a_{-1}$
and $a_{1}$ arbitrary. If we choose $a_{-1}=\beta \alpha ^{2}/\left(
\alpha ^{6}-1\right) $ and $a_{1}=\beta \alpha ^{4}/\left( \alpha
^{6}-1\right) $ then the choice $a_{-2}=\beta \alpha ^{4}/\left( \alpha
^{12}-1\right) $ yields $a_{K}=\beta K\alpha ^{4K}/\left( \alpha
^{6K}-1\right),$
which is clearly equivalent to $\widehat{f}_{1}\left( K\right)
$ when $\beta =6a_{2}$ and $\alpha=\exp \left( a_{2}/a_{1}\right) .$ Since all
continuous solutions
$\widehat{f}\left( k\right) $ give a corresponding solution
$a_{K}=\widehat{f}\left( K\right)$,
via (\ref{4d28}), we presume that other choices of $
a_{-2}$ give solutions for $a_{K}$ which are oscillatory and do not tend to
zero as $K\rightarrow \infty .$ As in the continuous case this gives
Fourier series which do not come from a continuous function of $x.$

It is also easy to show that there are additional solutions of (\ref{4d25}). If
the set $S_{1}=\left\{ a_{K}\right\} $ solves (\ref{4d26}) then so does the
set $ S=\left\{\gamma ^{K}a_{K}\right\} $ provided $\gamma ^{3K}=1$, for all
integer values of $K.$ This gives two additional solutions $\left\{
e^{2\pi Ki/3}a_{K}\right\} $ and $\left\{ e^{4\pi Ki/3}a_{K}\right\} .$ If
$f_{1}\left( x\right) $ is the $2\pi $-periodic function whose Fourier
series is given by the set $S_{1},$ then these two solutions are from the
functions $f_{1}\left( x+\frac{2}{3}\pi \right) $ and $f_{1}\left( x+\frac{4
}{3}\pi \right) $ that is the original function shifted by one third and
two thirds of its period respectively.

In this section we have shown, that by requiring $f$ to be bounded on
any compact subset of the real axis, we can recover all the solutions of
(1.5a,b) which have this boundedness property, by taking the the
appropriate Fourier Transform of (\ref{1d4}).  We have also shown that there
are no other solutions $f$ which have this boundedness property.
However by taking the appropriate limit we have also shown that the
$\delta $ function is also a solution and have recovered this from the
transformed equation. We have not obtained the $\wp $ function
solutions, which have double poles on the real axis. This is because the
transforms of these functions do not satisfy the transformed equation
(\ref{4d4}). The resolution of this problem is the topic of the next section.


\section{Even Solutions}

We now look at the solutions of (\ref{1d3}) which correspond to the even
solutions of (\ref{1d4}). Equation (\ref{4d4}) for the Fourier transform now
reads
\begin{equation}
\left( k\widehat{f}\left( l\right) +l\widehat{f}\left( k\right) \right) \widehat{f}
\left( k+l\right) =\left( k+l\right) \widehat{f}\left( k\right) \widehat{f}
\left(l\right) .
\label{5d1}
\end{equation}
Clearly if $\widehat{f}\left( k\right) $ is a function not identically zero then
(\ref{5d1}) implies that $k/\widehat{f}\left( k\right) $ is linear so that
$\widehat{f} \left( k\right) $ is constant. This however only gives the
distributional solution where $f\left( x\right) $ is a constant multiple of
$\delta \left( x\right)$, although as a distribution
$\widehat{f}\left( k\right) =a\delta \left(
k\right) $ is also a solution so that $\widehat{f}\left( k\right) =a\delta
\left( k\right) +b$ is a solution. These are the only even solutions
obtained via this method.

However, we know that the ${\wp }$ functions, which have double poles
at the origin, and in particular the function $1/x^{2}$ satisfy (\ref{1d3}). We
do not recover these solutions from (\ref{5d1}) because in these cases the left
hand side of equation (\ref{1d3}) is not identically zero but acts as a distribution.

For example, if we substitute $f=1/x^{2}$ into the left hand side of (\ref{1d3})
we get
\begin{eqnarray}
g\left( x_{1},x_{2},x_{3}\right) &=&-\dfrac{2}{\left( x_{1}-x_{2}\right)
^{3}\left( x_{1}-x_{3}\right) ^{2}}-\dfrac{2}{\left( x_{1}-x_{2}\right)
^{2}\left( x_{1}-x_{3}\right) ^{3}}  \notag \\
&&-\dfrac{2}{\left( x_{2}-x_{3}\right)^{3} \left( x_{2}-x_{1}\right) ^{2}} -
\dfrac{2}{\left( x_{2}-x_{3}\right) ^{2}\left( x_{2}-x_{1}\right) ^{3}}
\notag \\
&&-\dfrac{2}{\left( x_{3}-x_{1}\right) ^{3}\left( x_{3}-x_{1}\right) ^{2}}-
\dfrac{2}{\left( x_{3}-x_{1}\right) ^{2}\left( x_{3}-x_{2}\right) ^{3}}.
\label{5d2}
\end{eqnarray}
This expression is identically zero outside the neighbourhood of $
x_{1}-x_{2}=0,\;x_{1}-x_{3}=0,\;x_{2}-x_{3}=0.$ For fixed $x_{3}$ this is
the region outside the lines $x_{1}-x_{2}=0,$ $x_{1}=x_{3}$ and $x_{2}=x_{3}$
. In Appendix B we show that (\ref{5d2}) acts as a distribution supported only
at the point $x_{1}=x_{2}=x_{3}$ in the $(x_{1},x_{2})$ plane and can be
represented by the distribution
\begin{equation}
g\left( x_{1},x_{2},x_{3}\right) =\pi ^{2}\bigg( \delta ^{\prime }\left(
x_{1}-x_{3}\right) \delta ^{\prime \prime }\left( x_{2}-x_{3}\right) +\delta
^{\prime \prime }\left( x_{1}-x_{3}\right) \delta ^{\prime }\left(
x_{2}-x_{3}\right) \bigg) .
\label{5d3}
\end{equation}
Hence if we take the Fourier transform of (\ref{1d4}) assuming that $f$ is even
and has a double pole at the origin but is otherwise bounded, we have
instead of $\widehat F =0,$
\begin{eqnarray}
\widehat F &=&\iint\limits_{\mathbb{R}\sp{2}}\pi ^{2}\big( \delta ^{\prime
}\left( x_{1}-x_{3}\right) \delta ^{\prime \prime }\left( x_{2}-x_{3}\right)
+\delta ^{\prime \prime }\left( x_{1}-x_{3}\right) \delta ^{\prime }\left(
x_{2}-x_{3}\right) \big) xe^{-ikx_{1}-ilx_{2}}dx_{1}dx_{2} \notag\\
&=&i\pi ^{2}kl\left( k+l\right) e^{-i\left( k+l\right) x_{3}}.
\label{5d4}
\end{eqnarray}
Thus (\ref{4d4}) becomes
\begin{equation}
\left( k\widehat{f}\left( l\right) +l\widehat{f}\left( k\right) \right) \widehat{f}
\left( k+l\right) =\left( k+l\right) \widehat{f}\left( k\right) \widehat{f}\left(
l\right) +\pi ^{2}kl\left( k+l\right) .
\label{5d5}
\end{equation}
The Fourier transform of $1/x^{2}$ is $-\pi \left| k\right| $ and the
identity
\begin{equation}
\left( k\left| l\right| +l\left| k\right| \right) \left| k+l\right| =\left(
k+l\right) \left( \left| k\right| \left| l\right| +kl\right) ,
\label{5d6}
\end{equation}
for all $k,l$ ensures that the even functions $\widehat{f}\left( k\right) =
\pm \pi \left| k\right| $ satisfy (\ref{5d5}). We note here that if
$\widehat{f}(k)$ satisfies (\ref{5d5}) so does $-\widehat{f}\left( k\right)$.
However, in addition to solving (\ref{5d5}) we also require that the inverse $
f\left( x\right) $ satisfies the condition that $x^{2}f\left( x\right)
\rightarrow +1$ as $x\rightarrow 0.$ Thus although $\widehat{f}\left( k\right)
=\pm \pi \left| k\right| $ satisfies (\ref{5d5}) we must take the solution
$\widehat{f}\left( k\right) =-\pi \left| k\right|$. However,
$\widehat{f}\left( k\right) =\pm \pi \left| k\right| $ are not the only
solutions of (\ref{5d5}). Writing (\ref{5d5}) as
\begin{equation}
\widehat{f}(l) \left( k\left(\widehat{f}(k+l)
-\widehat{f}(k) \right) -l\widehat{f}(k)\right)
+l\widehat{f}\left( k\right) \widehat{f}
\left( k+l\right) =\pi ^{2}kl\left( k+l\right)
\end{equation}
and then dividing by $l$ and taking the limit as $l\rightarrow 0$ gives
\begin{equation}
a_{0}k\widehat{f}^{\prime }\left( k\right)
-a_{0}\widehat{f}\left( k\right) +\widehat{f}^{2}\left( k\right)=\pi^{2}k^{2},
\label{5d8}
\end{equation}
where $a_{0}=\widehat{f}\left( 0\right) $ and the solution of (\ref{5d8}) with
$\widehat{f} \left( 0\right) =a_{0}$ is
\begin{equation}
\widehat{f}\left( k\right) =\pi k\coth \left( \pi k/a_{0}\right) .
\label{5d9}
\end{equation}
This is the Fourier transform of the function
\begin{equation}
f\left( x\right) =-\dfrac{1}{4}a_{0}\left| a_{0}\right| /\sinh^{2}\left(
\dfrac{a_{0}x}{2}\right) .
\end{equation}
Again, apart from the addition of a constant and a scaling of $x$ this
produces the unique ${\wp }$ function of imaginary period $\pi $ and
real period infinity.

For functions $f\left( x\right) $ which are $2\pi $-periodic in addition to
having a double pole of the form $1/x^{2}$  at the origin,  we need to modify
(\ref{5d3}) and (\ref{5d4}). The distribution
$g\left( x_{1},x_{2},x_{3}\right) $ must be replaced by one which is
repeated $2\pi $ periodically in $x_{1}$ and $x_{2}$ that is
\begin{equation}
g_{1}\left( x_{1},x_{2},x_{3}\right) =\sum\limits_{p,q}C_{pq}e^{i\left(
p\left( x_{1}-x_{3}\right) +q\left( x_{2}-x_{3}\right) \right) },
\end{equation}
so that
\begin{equation}
\widehat g = 4\pi ^{2}
\sum\limits_{p,q}C_{pq}\delta \left( k-p\right) \delta \left( l-q\right)
e^{-i\left( p+q\right) x_{3}} ,
\end{equation}
where
\begin{equation}
C_{pq}=\dfrac{1}{4}pq(p+q)i.
\end{equation}
Equation (\ref{5d5}) now becomes
\begin{equation}
\left( k\widehat{f}(l)+l\widehat{f}\left( k\right) \right) \widehat{f}\left( k+l\right)
=\left( k+l\right) \widehat{f}\left( k\right) \widehat{f}\left( l\right)
+\sum_{K,L}\pi ^{2}KL\left( K+l\right) \delta \left( k-K\right)
\delta \left( l-L\right) .
\end{equation}
Again following (\ref{4d24}) we write $\widehat{f}\left( k\right) $ in the form
\begin{equation}
\widehat{f}\left( k\right) =\sum\limits_{p=-\infty }^{\infty }a_{p}\delta
\left( k-p\right)
\end{equation}
and obtain the following recurrence relation for the set $\left\{
a_{K}\right\} $
\begin{equation}
\left( Ka_{L}+La_{K}\right) a_{K+L}=\left( K+L\right) a_{K}a_{L}+\pi
^{2}KL(K+L).
\end{equation}

As in \S4 with the recurrence relation (\ref{4d25}), $a_{0}$ is arbitrary. \
Writing $a_{K}=\pi Kb_{K}$ for $K=1,2,...$ we obtain
\begin{equation}
\left( b_{2}+b_{K}\right) b_{K+L}=b_{K}b_{L}+1
\end{equation}
and with $b_{1}=\dfrac{\beta +1}{1-\beta }$ and $L=1$ we have,
\begin{equation}
b_{K+1}=\left( b_{K}\left( \beta +1\right) +1-\beta \right) /\left(
b_{K}\left( 1-\beta \right) +\beta +1\right).
\end{equation}
This is easily solved to get
\begin{equation}
b_{K}=\dfrac{1+\beta ^{K}}{1-\beta ^{K}}\;\;\text{or\ }a_{K}=\pi \left|
K\right| \left( \dfrac{1+\beta ^{\left| K\right| }}{1-\beta ^{\left|
K\right| }}\right) .
\label{5d19}
\end{equation}
This reproduces the result that $a_{K}=\widehat{f}\left( K\right) $ at integer
values of $K$ where $\widehat{f}\left( K\right) $ is the continuous version.
Also the Fourier series $\sum\limits_{-\infty }^{\infty }\pi \left| K\right|
\left( \dfrac{1+q^{2\left| K\right| }}{1-q^{2\left| K\right| }}\right)
e^{iKx}$ can be recognised as the Fourier series for the ${\wp }$
functions again up to the addition of a constant. Here $q=\beta ^{1/2}$ is
the usual notation for the nome.

The conclusion is that the only even solution of (\ref{1d4}) are those of (1.5b)
with $d=0.$


\section{The General Case}

We now consider the even solutions of (\ref{eqgen}) for a general integer $n.$
It will be convenient to  write (\ref{eqgen}) as
\begin{equation}
g\left( \mathbf{x},x_{n+1}\right) =\sum\limits_{p=1}^{n+1}\dfrac{\partial }{
\partial x_{p}}\prod\limits_{q \neq p} f\left( x_{p}-x_{q}\right) ,
\end{equation}
where $f$ is even and $\mathbf{x}$ is the vector $(x_{1},x_{2}, \cdots
x_{n})$. We then define the n-dimensional Fourier transform
$\widehat g( \mathbf{k},x_{n+1}))\equiv T[g] ( \mathbf{k},x_{n+1})$ of
$g\left( \mathbf{x},x_{n+1}\right) $ by
\begin{equation}
\widehat g\left( \mathbf{k},x_{n+1}\right) =\int_{\mathbb{R}^{n}}g\left(
\mathbf{x,}x_{n+1}\right) e^{-i\mathbf{k.x}}d\mathbf{x,}
\label{6d2}
\end{equation}
as a generalisation of (\ref{4d1}). Again there are problems with the double
pole of $f$ at the origin. Now however we find that $g$ acts as a
distribution over any plane through the origin and the singularities become
more difficult to deal with. To overcome this difficulty we assume that
the arguments of $g$ are complex and we replace $x_{j}$ by $
x_{j}+i\epsilon_{j}$ where $\epsilon_{n+1}>\epsilon_{n}...>\epsilon_{1}>0$
and assume that $\epsilon_{n+1}$ is small. In other words
\begin{equation}
g=\sum\limits_{p=1}^{n+1}\dfrac{\partial }{\partial x_p}\prod\limits_{q \neq
p}f\left( x_{p}-x_{q}+i\left( \epsilon_{p}-\epsilon_{q}\right) \right) .
\label{6d3}
\end{equation}
We then assume that in the definition of $\widehat g$, (\ref{6d2}),
we integrate along the Real axis in the complex $x_{j}+iy_{j}$ plane. If the
function $f\left(
z\right) $ has double poles on the $x=Re\;z$ axis and no other singularities
in the neighbourhood of the $x$-axis then provided $\epsilon_{n+1}$ is small
enough, there will be no singularities of $g$ within the domain of
integration of the integral occurring in (\ref{6d2}).

To calculate $\widehat g $ we first write  $\xi_q= x_{1}-x_{n+1}+i\left(
\epsilon _{1}-\epsilon _{n+1}\right)$ when $q=1,$ $\xi_q= x_{1}-x_{q}+i\left(
\epsilon _{1}-\epsilon _{q}\right)$ when
$2\leq q \leq n+1,$ and  $\eta_q= X_{q}+i\left(
\epsilon _{1}-\epsilon _{q}\right)$,
where $X_{q}=x_1-x_q$. Noting that $\xi_1=\xi_{n+1}$ we  consider
\begin{eqnarray}
I &=&\int\limits_{x_{1}=-\infty }^{\infty
}\int\limits_{\mathbb{R}^{m-1}}\prod\limits_{q=2}^{n+1}f\left(\xi_q
\right) e^{-i\mathbf{k.x}}dx^{m-1}dx_{1}
\notag \\
&=&\int\limits_{x_{1}=-\infty }^{\infty }f(\xi_{1})e^{-ik_{1}x_{1}}\left(
\prod\limits_{q=2}^{n}\ \ \int\limits_{x_{q}=-\infty }^{\infty }f\left(\xi_q
\right)e^{-ik_{q}x_{q}}dx_{q}\right) dx_{1}  \notag \\
&=&\int\limits_{x_{1}=-\infty }^{\infty }f(\xi_{1})e^{-ik_{1}x_{1}}\left(
\prod\limits_{q=2}^{n}\ \ \int\limits_{X_{q}=-\infty }^{\infty }f\left(
\eta_{q} \right)
e^{ikq(X_{q}-x_{1})}dX_{q}\right) dx_{1}.
\label{6d4}
\end{eqnarray}
Now since $\epsilon _{q}>\epsilon _{1}$ the singularities of $f\left(
z_{1}\right) $, at $z_{1}=i(
\epsilon _{n+1}-\epsilon _{1})$ and
$f\left(
z_{q}\right) $, at $z_{q}=i(
\epsilon _{q}-\epsilon _{1})$, $q>1$, lie in the upper half plane.

For functions $f\left(z\right) $ which are analytic in the neighbourhood of
the $Re\;z$ axis but have a pole at $z=0$ we define Fourier transforms
$\widehat{f}_{U}$ and $\widehat{f}_{L}$ by
\begin{equation}
\widehat{f}_{U}\left( k\right) =\int\limits_{-\infty }^{\infty
}\!\!\!\!\!\!\!\cap f\left( x\right) e^{-ikx}dx
\label{6d5}
\end{equation}
and
\begin{equation}
\widehat{f}_{L}\left( k\right) =\int\limits_{-\infty }^{\infty
}\!\!\!\!\!\!\!\cup f\left( x\right) e^{-ikx}dx,
\end{equation}
which are suitably indented to go above and below the singularity at the
origin. With these definitions we have, in the limit as
$\epsilon_{n+1}\rightarrow 0$,
\begin{eqnarray}
I &=&\int_{-\infty }^{\infty } f\left( x_{1}-x_{n+1}+i\left( \epsilon
_{1}-\epsilon _{n+1}\right) \right) e^{-i\left(
\sum\limits_{q=1}^{n}k_{q}\right) x_{1}}
dx_{1}\prod\limits_{q=2}^{n}\widehat{f}_{L}\left( -k_{q}\right)   \notag \\
&=&e^{-i\left( \sum\limits_{q=1}^{n}k_{q}\right) x_{n+1}}\widehat{f}_{L}\left(
\sum\limits_{q=1}^{n}k_{q}\right) \prod\limits_{q=2}^{n}\widehat{f}_{L}\left(
-k_{q}\right) .
\end{eqnarray}
Thus the first term in the sum in (\ref{6d3}) contributes a term $ik_{1}I$ to
$\widehat g.$ A similar calculation gives
\begin{multline}
\widehat g  =\left( \sum\limits_{j=1}^{n}\left[ k_{j}\left(
\prod\limits_{q=1}^{j-1}\widehat{f}_{U}\left( -k_q\right) \right) \left(
\prod\limits_{q=j+1}^{n}\widehat{f}_{L}\left( -k_{q}\right) \right) \right]
\widehat{ f}_{L}\left( \sum\limits_{q=1}^{n}k_{q}\right) \right. \\
\left. -\left( \sum_{j=1}^{n}k_{j}\right) \left( \prod\limits_{q=1}^{n}
\widehat{f}_{U}\left( -k_{q}\right) \right) \right)
ie^{-i\left( \sum\limits_{q=1}^{n}k_{q}\right) x_{n+1}}
\end{multline}
so that $\widehat g =0$ gives
\begin{equation}
\sum\limits_{j=1}^{n}\left[ k_{i}\left( \prod\limits_{q=1}^{j-1}\widehat{f}
_{U}\left( -k_{q}\right) \right) \left( \prod\limits_{q=j+1}^{n}\widehat{f}
_{L}\left( -k_{q}\right) \right) \right] \widehat{f}_{L}\left(
\sum\limits_{q=1}^{n}k_{q}\right)
-\left( \sum\limits_{j=1}^{n}k_{j}\right)
\left( \prod\limits_{q=1}^{n}\widehat{f}_{U}\left( -k_{q}\right) \right) =0\;.
\label{6d9}
\end{equation}

We now define the Fourier transform of $f\left( z\right) $ to be
$\frac{1}{2}\left( \widehat{f}_{U}\left( k\right) +
\widehat{f}_{L}\left( k\right) \right)$.
This corresponds to the usual generalised definition of the Fourier
transform of functions with non-integrable singularities. This also
coincides with the use in \S5. Also we have the result that the difference
$\widehat{f}_{L}\left( k\right) -\widehat{f}_{U}\left( k\right) $ is $
2\pi i$ multiplied by the residue of the functions $f\left( z\right) e^{-ikz}$
at the origin, assuming that the only singularity of $f\left( z\right) $ on
the real axis is at the origin. We also make the assumption that $f\left(
z\right) $ is an even function of $z$ with a double pole of the form $1/z^{2}
$ at the origin. Hence with
\begin{equation}
\widehat{f}\left( k\right) =\dfrac{1}{2}\left( \widehat{f}_{U}\left( k\right) +\widehat{f
}_{L}\left( k\right) \right)
\label{6d10}
\end{equation}
and
\begin{equation}
\widehat{f}_{L}\left( k\right) -\widehat{f}_{U}\left( k\right) =2\pi i\times \text{
Residue}\left( f\left( z\right) e^{-ikz}\right) =2\pi k,
\end{equation}
we have
\begin{equation}
\widehat{f}_{L}\left( k\right) =\widehat{f}(k)+\pi k
\end{equation}
and
\begin{equation}
\widehat{f}_{U}\left( k\right) =\widehat{f}\left( k\right) -\pi k.
\end{equation}

Hence the equation $\widehat g =0,$ (\ref{6d9}) gives
\begin{multline}
S_{n} \equiv
\sum\limits_{j=1}^{n}\left[ k_{j}\prod\limits_{q=1}^{j-1}\left( \widehat{f}
\left( k_{q}\right) +\pi k_{q}\right) \prod\limits_{q=j+1}^{n}\left( \widehat{f}
\left( k_{q}\right) -\pi k_{q}\right) \right] \left( \widehat{f}\left(
\sum\limits_{q=1}^{n}k_{q}\right) +\pi \sum\limits_{q=1}^{n}k_{q}\right)
 \\
-\left( \sum\limits_{j=1}^{n}k_{j}\right) \left(
\prod\limits_{q=1}^{n}\left( \widehat{f}\left( k_{q}\right) +\pi k_{q}\right)
\right) =0.
\label{6d14}
\end{multline}
This equation is to be regarded as an equation which determines $\widehat{f}
\left( k\right) $ and must be satisfied for all $\left\{ k_{q}\right\} $ in $
\mathbb{R}^{n}.$ Again we note that, by inspection, $\widehat{f}\left( k\right)
=\pm \pi \left| k\right| $ satisfies (\ref{6d14}) and we may also show that
$\widehat{f} (k)=t\pi |k|$ satisfies (\ref{6d4}) for the all values of $t$ which
satisfy $\left( 1+t\right) ^{n}\left( t-1\right) =\left( t-1\right)
^{n}\left( 1+t\right).$ Apart from $t=\pm 1$, these are,  $t=i$
$\cot \dfrac{j\pi}{n-1},\;\; j=1,\ldots n-2$.
However, to satisfy the requirement that the inverse $f\left( z\right) $
is such that $z^{2}f\left( z\right) \rightarrow 1$ as $z\rightarrow 0$ we
require the solution above with $t=-1.$ This problem recurs throughout
\S6 and \S 7 and we will assume that we only take the multiple of
$\widehat{f}\left( k\right) $ which satisfies the criterion that it has the
correct double pole either at the origin or at the sequence of double
poles when $f\left( z\right) $ is periodic. With $m=2$ it is a simple
matter to verify that $S_{2}=0$ is equivalent to (\ref{5d5}).

The following lemmas prove useful in finding the complete solution to
(\ref{6d14}) for all $n$. We begin with a definition.
\begin{defn}
For each $n$ let\newline
\indent $\mathcal{F}_{n}$ be the solution set of  $S_{n}=0$,\newline
\indent $\mathcal{G}_{n}$ be the solution set of  $S_{n}=0$ and
$\widehat{f}\left(0\right) \neq 0$.
\end{defn}

\begin{lem}
For $n \geq 4$  and even, $\mathcal{F}_{n}\subseteq \mathcal{F}_{2}.$
\end{lem}

\begin{lem}For $n\geq 3,\;\;\;\mathcal{G}_{n}\subseteq \mathcal{G}_{2}.$
\end{lem}

\noindent{\bf Proof of Lemma 1:}
In $S_{n+2}$ we put $k_{n+2}=-k_{n+1}$ and using the fact that $\widehat{f
}$ is even we find
\begin{equation}
S_{n+2}\left| _{_{_{k_{n+2}=-k_{n+1}}}}=\left( \widehat{f}^2\left( k_{n+1}
\right) -\pi ^{2}k_{n+1}^{2}\right) S_{n}.\right.
\label{6d15}
\end{equation}
The factor $\widehat{f}^{2}\left( k\right) -\pi ^{2}k^{2}$ produces the
solutions $\widehat{f}\left( k\right) =\pm \pi \left| k\right| $ if
$\widehat{f}$ is even. This function we have already shown (see (\ref{5d5}),
(\ref{5d6})) belongs to the solution set $\mathcal{F}_{2}$. Since the solution
set $\mathcal{F}_{n+2}$ must be contained in the solution set of
$S_{n+2}\left| _{k_{n+2}=1-k_{n+1}},\right. $ (\ref{6d15})
shows that $\mathcal{F}_{n+2}\subseteq \left( \widehat{f}\left( k\right) =
\pm \left| k\right| \right) \cup \mathcal{F}_{n}$.
The result now follows by induction.

\noindent{\bf Proof of Lemma 2:}
In $S_{n+1}$ we put $k_{n+1}=0$ and obtain
\begin{equation}
S_{n+1}\left| _{_{_{k_{n+1}=0}}}=\widehat{f}(0)S_{n}.\right.
\end{equation}
Again if $\widehat{f}(0)\neq 0$ the result follows by induction.

\begin{defn} Let $\mathcal{B}$ denote the class of functions
whose generalised Fourier transform
$\widehat{f}\left( k\right) $ arise from functions $f(z)$ bounded on the
real axis apart from a double pole of the form $1/z^{2}$ at the origin.
\end{defn}
This is the class of functions for which we have derived (\ref{6d14}). We then
have
\begin{thm}
For solutions with $\widehat{f}\in \mathcal{B}$, then
$\mathcal{F}_{n}=\mathcal{F}_{2}$ for all even $n$.
\label{th2}
\end{thm}

\noindent{\bf Proof:}  By Lemma 1 $\mathcal{F}_{n}\subseteq \mathcal{F}_{2}$.
In \S5 we have proved that $\mathcal{F}_{2}$ is the one parameter family
$\widehat{f}\left( k\right) =\pi
k\;\coth \left( \pi k/a_{0}\right) ,$ together with its limit as $
a_{0}\rightarrow 0,\;\pm \pi \left| k\right| .$ It is easily verified by
substitution that these satisfy $S_{n}=0.$ This gives
$\mathcal{F}_{2}\subseteq \mathcal{F}_{n}$. The result now follows.

\begin{thm}
For solutions with $\widehat{f}\in \mathcal{B}$, then
$\mathcal{G}_{n}=\mathcal{G}_{2}$ for all  $n$.
\label{th3}
\end{thm}

The proof of theorem \ref{th3} follows by the same method as the  proof of
theorem \ref{th2}.  However we cannot exclude the possibility that there exist
solutions of $S_{n}=0,$ with $n \geq 3$ and  odd, which have
$\widehat{f}\left(0\right) =0,$ but are not contained in the solutions of
$S_{2}=0$. Before we look at the
possibility of such solutions we consider the extensions of Lemma 1 and 2
and the corresponding theorems about the solution sets, to functions $
f\left( z\right) $ which are periodic. This requires $f\left( z\right) $
to have a periodic array of double poles of the form ${1}/{\left(
z-2n\pi \right) ^{2}}$ at the points $z=2n\pi$, $n\in \mathbb{Z}$.
Again we follow the method outlined in \S4 and write $\widehat{f}
\left( k\right) $ as in (\ref{4d24}) and reproduce from $S_{n}=0$ a similar
equation with $\widehat{f}\left( k_{q}\right) $ replaced by $a_{k_{q}}$, which
is equivalent to (\ref{4d26}).
The conclusions can be summed up in the following theorem, which incorporates
the result of Braden and Byatt-Smith \cite{BBS} for even functions
as a special case.
\begin{defn} Let ${\mathcal{B}}_{2\pi}$ denote the class of functions
whose generalised Fourier transform $\widehat{f}\left( k\right) $ arise from 
$2\pi$ periodic functions $f(z)$ bounded on the real axis apart from 
double poles at $1/(z-2\pi n)^{2}$, $n\in\mathbb{Z}$.
\end{defn}

\begin{thm} For $\widehat{f}\in{\mathcal{B}}\cup{\mathcal{B}}_{2\pi}$\newline
1. The only even solutions of (\ref{eqgen}) with
$n$ even are those of (1.5b) with $d=0$; \newline
2. The only even solutions of (\ref{eqgen}) with $n$ odd and
for which $\widehat{f}\left( 0\right) \neq 0$ are those of (1.5b) with $d=0$.
\end{thm}


\section{New Solutions}

We now look at the solution of (\ref{eqgen}) which have
$\widehat{f}\left( 0\right) =0$, to see if there are solutions which do not
belong to the set $\mathcal{F}_{2}$.
Lemma 1 and Theorem \ref{th2} can easily be adapted to prove that when $n$ is
odd $\mathcal{F}_{n}\subseteq \mathcal{F}_{3}.$ We first find this solution set
and then prove that when $n$ is odd $\mathcal{F}_{n}=\mathcal{F}_{3}$.
Hence we look for solutions of (\ref{6d14}), with
$\widehat{f}\left( 0\right) =0,$ when $n=3$. We wish to consider only even
functions, but wish to include functions like $\left| k\right| $ which is
not differentiable at $k=0$. Hence we consider (\ref{6d14}) defined on the
subspace $k_{q}\geq 0,$ $q=1,2$ and $3,$ with all derivatives at the origin
defined by one sided derivatives. Thus in the interval $k\geq 0,\;\left|
k\right| $ is defined as $k$ with derivative $1$ at the origin.

In (\ref{6d14}) we write $k_{1}=k$ and $k_{2}=k_{3}=l$ and let
$l\rightarrow 0$. Then (\ref{6d14}) gives
\begin{equation}
a_{0}\left( a_{0}k\widehat{f}^{\prime }\left( k\right) -a_{0}\widehat{f}\left( k\right) +
\widehat{f}^{2}\left( k\right) -\pi ^{2}k^{2}\right) =0,
\end{equation}
where $a_{0}=\widehat{f}\left( 0\right) .$ When $a_{0}\neq 0$ this gives the
same equation as (\ref{5d8}) but is automatically satisfied when $a_{0}=0.$ So
in addition to the solution given in (\ref{5d9}), which belongs to
$\mathcal{F}_{2},$ we can also allow $a_{0}=0$.

When $a_{0}=0$ the next term in the expansion of (\ref{6d14}) gives
\begin{equation}
a_{1}\left( \widehat{f}^{2}\left( k\right) -\pi ^{2}k^{2}\right) =0,
\label{7d2}
\end{equation}
where $a_{1}=\widehat{f}^{\prime }\left( 0\right)$. If $a_{1}\neq 0$ then
(\ref{7d2})
gives $\widehat{f}\left( k\right)= \pm \pi \left| k\right| $ as the only
even solution. This is also the solution of (\ref{5d9}) when $a_{0}=0$ and also
belongs to $\mathcal{F}_{2}$. Now if we assume that $a_{1}=0,$ we can write the
third term in the expansion of (\ref{6d14}) as
\begin{equation}
2\pi ^{2}k\widehat{f}^{\prime }\left( k\right) +a_{2}\widehat{f}^{2}\left(
k\right) -2\pi ^{2}\widehat{f}\left( k\right) =\pi ^{2}a_{2}k^{2},
\end{equation}
where $a_{2}=\widehat{f}^{\prime \prime }\left( 0\right) .$ The only even
solution of this equation is
\begin{equation}
\widehat{f}\left( k\right) =\pi k\tanh \left(\dfrac{ka_{2}}{2\pi} \right) ,
\label{7d4}
\end{equation}
which automatically has $f^{\prime \prime }\left( 0\right) =a_{2}$. This of
course is a necessary requirement and we need to check that this is a
solution of (\ref{6d14}).

We rewrite (\ref{6d14}) as
\begin{equation}
\widehat{S}_{n} \equiv \left( \sum\limits_{j=1}^{n}\dfrac{k_{j}}{\widehat{f}\left(
k_{j}\right) +\pi k_{j}}\prod\limits_{q=j+1}^{n}\left( \dfrac{\widehat{f}\left(
k_{q}\right) -\pi k_{q}}{\widehat{f}\left( k_{q}\right) +\pi k_{q}}\right)
\right)
\dfrac{\left( \widehat{f}\left( \sum\limits_{j=1}^{n}k_{j}\right) +\pi
\sum\limits_{j=1}^{n}k_{j}\right) }{\sum\limits_{j=1}^{n}k_{j}}-1
=0.
\label{7d5}
\end{equation}
Substituting $\widehat{f}\left( k\right) =\pi k\tanh \left( ak\right) $
into $\widehat{S}_{n}$ gives
\begin{equation}
\widehat{S}_{n}=\left( \sum\limits_{j=1}^{n}\left( e^{2ak_{j}}+1\right)
\prod\limits_{q=j}^{n}\left( -e^{-2ak_{q}}\right) \right) \dfrac{\exp \left(
2a\sum\limits_{j=1}^{n}k_{j}\right) }{\exp \left(
2a\sum\limits_{j=1}^{n}k_{j}\right) +1}-1.
\label{7d6}
\end{equation}
The first term can be written as
\begin{equation}
\widehat{S}_{n}^{(1)}=\sum\limits_{j=1}^{n}a_{j+1}-a_{j}\equiv a_{n+1}-a_{1},
\end{equation}
where $a_{j}= \prod\limits_{q=j}^{n}\left( -e^{-2ak_{q}}\right) $ so
that $\widehat{S}_{n}^{(1)}=\left( -1\right) ^{n+1}\exp \left(
-2a\sum\limits_{j=1}^{n}k_{q}\right) +1.$ Hence
\begin{equation}
\widehat{S}_{n}=\dfrac{\left( \left( -1\right) ^{n+1}-1\right) }{\exp \left(
2a\sum\limits_{j=1}^{n}k_{j}\right) +1}.
\label{7d8}
\end{equation}
This gives immediately $\widehat{S}_{n}\equiv 0$ whenever $n$ is odd. Hence $
\widehat{f}\left( k\right) =\pi k\tanh \left( ak\right) ,$ with $a$ arbitrary,
is a solution for all equations $S_{n}=0$, when $n$ is odd. It is also evident
from (\ref{7d8}) that this solution does not satisfy $S_{n}=0$ for $n$ even.

We also note that substituting $\widehat{f}\left( k\right) =\pi k\coth \left(
ak\right) $ into (\ref{6d5}) changes (\ref{7d6}) to
\begin{equation}
\widehat{S}_{n}=\left( \sum\limits_{j=1}^{n}\left( e^{2ak_{j}}-1\right)
\prod\limits_{q=j}^{n}e^{-2ak_{q}}\right) \dfrac{\exp
\sum\limits_{j=1}^{n}\left( 2ak_{j}\right) }{\exp \left(
\sum\limits_{j=1}^{n}2ak_{j}\right) -1}-1.
\end{equation}
The change in signs now means that $\widehat{S}_{n}\equiv 0$ for all $n$ showing
that $\widehat{f}_{1}\left( k\right) =\pi k\coth \left( ak\right)$,  $a$
arbitrary satisfies (\ref{6d14}) for all $n$ as indicated earlier.

If we write $a=\frac{1}{2}\pi /\alpha $ so that $\widehat{f}\left( k\right)
=\pi k\tanh \left( \frac{1}{2}\pi k/\alpha \right) $ then this is the
Fourier transform of the function $f\left( z\right) =-\alpha |\alpha |{
\cosh \alpha z}/{\sinh ^{2}\alpha z},$ so $\alpha $ must be negative to
satisfy the pole condition that $z^{2}f\left( z\right) \rightarrow 1$ as $
z\rightarrow 0.$ However, since the coefficient of the double pole is in
fact arbitrary we have $f\left( z\right) =\beta {\cosh \alpha z}/{\sinh
^{2}\alpha z}$ satisfies (\ref{eqgen}) for all odd values of $n$.
 This is the solution referred to in \S2.

The even period solutions, which satisfy the modification of (\ref{6d14}) when $
f\left( z\right) $ has an array of double poles of the form $\left( z-2p\pi
\right) ^{-2}$ at the points $z=p\pi ,$ $p=0,\pm 1,\pm 2,...,$ can be
written as
\begin{equation}
\widehat{f}\left( k\right) =\sum\limits_{p=-\infty }^{\infty } a_{p}\delta
\left( k-p\right) ,\;\;\text{with\ }a_{-p}=a_{p},
\end{equation}
as in (\ref{4d24}). Again, (\ref{6d14}) is now to be satisfied at all
integer values
of $\left\{ k_{q}\right\} $ with $\widehat{f}\left( k_{q}\right) $ replaced by $
a_{k_{q}}.$ Hence
\begin{multline}
\widehat {S}_{n}\equiv \sum\limits_{j=1}^{n}\left[ N_{j}\prod\limits_{q=1}^{j=1}\left(
a_{N_{q}}+\pi N_{q}\right) \prod\limits_{q=j+1}^{n}\left(
a_{N_{q}}-\pi N_{q}\right) \right]   \\
\left( a_{\sum\limits_{q=1}^{n}N_{q}}+\pi \sum\limits_{q=1}^{n}N_{q}\right)
-\left( \sum\limits_{j=1}^{n}N_{j}\right) \left(
\prod\limits_{q=1}^{n}\left( a_{N_{q}}+\pi N_{q}\right)\right) =0.
\end{multline}
To solve $\widehat {S}_{3}=0$ we proceed as in \S5. If $a_{0}$ is not
equal to zero we can put $N_{3}=0$ and recover the solutions (\ref{5d19}). \
However, if $a_{0}=0$ then $\widehat {S}\equiv 0$ for all $N_{1}$ and $N_{2}$
if $N_{3}=0.$ Writing down all the equations for $N_{j}\geq 1$ it is easy
to establish that if $a_{1}=\pi \left( \beta +1\right) /\left( 1-\beta
\right) $ the odd terms are given by
\begin{equation}
a_{2N+1}=\pi \left( 2N+1\right) \dfrac{\left( 1+\beta ^{2N+1}\right) }{
1-\beta ^{2N+1}},\;N\geq 0,
\end{equation}
as in (\ref{5d19}) but the even terms depend on the choice of $a_{2}$ which must
take one of the values
\begin{equation}
a_{2}=2\pi \dfrac{\left( 1+\beta ^{2}\right) }{1-\beta ^{2}}\;\;\text{or\ }
2\pi \dfrac{\left( 1-\beta ^{2}\right) }{1+\beta ^{2}}.
\end{equation}
The first choice gives
\begin{equation}
a_{2N}=2\pi N\dfrac{\left( 1+\beta ^{2N}\right) }{1-\beta ^{2N}},\;\;N\geq 0,
\label{7d14}
\end{equation}
and the second
\begin{equation}
a_{2N}=2\pi N\dfrac{\left( 1-\beta ^{2N}\right) }{1+\beta ^{2N}},\;\;N\geq 0.
\label{7d15}
\end{equation}

The result (7.13a) and (\ref{7d14}) is equivalent to (\ref{5d19}) while
writing $\beta =-\tilde{\beta}$ (7.13b) and (\ref{7d15}) is equivalent to
\begin{equation}
a_{N}=\pi \left| N\right| \left( \dfrac{1-\beta ^{\left| N\right| }}{1+\beta
^{\left| N\right| }}\right) ,\;\;\text{for all }N.
\end{equation}
This reproduces the result of (\ref{7d4}) at integer values of $k.$ The proof
that $\left\{ a_{N}\right\} $ satisfies $\widehat {S}_{n}$ for all $n$ is
identical to the proof in the continuous case, (see (\ref{7d5}) - (\ref{7d8})).

The corresponding inverse $f\left( z\right)$, obtained from $\widehat{f}(k)$
can be constructed either by taking the Fourier inverse of $\widehat{f}\left(
k\right) $ or by the infinite sum defined by (3.27) using the function $
\beta \cosh \alpha z/\sinh ^{2}\alpha z.$ As noted in \S3,
this function must be one of cn/sn$^{2}$,\;dn/sn$^{2}$ or
cndn/sn$^{2}.$ There are two reasons why there are three functions
representing the solution set. The first is that if all the parameters
defining the elliptic function are real then the transformation $
z\rightarrow iz$ permutes these three functions according to Jacobi's
Imaginary transformation sn$\left( iz,k\right) \rightarrow i$ sn$\left(
z,1-k^{2}\right) /$cn$\left( z,1-k^{2}\right) ,$ cn$\left( iz,k\right)
\rightarrow 1/$cn$\left( z,1-k^{2}\right) $ and dn$\left( iz,k\right)
\rightarrow $dn$\left( z,1-k^{2}\right) /$cn$\left( z,1-k^{2}\right) .$ \
Secondly, Jacobi's Real transformation defines the elliptic function for the
parameter $k<0$ or $k>1$ in terms of elliptic functions with a scaled
independent variable and parameter $k$ in the range $0<k<1.$ Again the
effect is to permute the three functions.

The conclusion is that the even solutions of (\ref{1d4}) with $n$ odd fall into
two categories. The set defined by (1.5b) with $d=0,$ which satisfy (\ref{1d8})
subject to $z^2f\left( z\right) \rightarrow $ a constant as $z\rightarrow 0$
and also the set which satisfy (A1.4) subject to the same condition at the
origin. These solutions have been expressed in terms of the ${\wp }$
function and also in terms of the Jacobian elliptic functions and
theorem 1 is proved.

\section{Acknowledgements}
We wish to thank A. M. Davie for helpful discussion and sharing his insight.
One of the authors (H.W.B.) wishes to thank the Newton Institute for 
support during the completion of this work.
\pagebreak
\appendix
\renewcommand{\thesection}{Appendix~\Alph{section} }
\renewcommand{\theequation}{A1.\arabic{equation}}
\section{}
{\setlength{\textwidth}{12 true cm}
\enlargethispage{11in}
The expansion procedure, for $n=3,$ described in \S 3 yields the
following two simultaneous equations (A1.1) and (A1.2) for the function
$f\left( x\right)$, which satisfies (\ref{eqgen})
\begin{equation}
120\,a_2\,{f^{\prime }}\,f + 5\,{a_{0}}\,
{f^{\prime \prime \prime }}\,{f^{\prime \prime }} + 60\,{f^{\prime \prime }}\,{f^{\prime }}\,{a_1}
 - {a_{0}}\,{f^{(5)}}\,f=0
\end{equation}
and

$ 504\,{a_3}\,{f^{\prime }}\,f^{2} + 1080\,
{f^{\prime }}\,a_2\,{f^{\prime \prime }}\,f + 36\,{{f^{(iv)}}}\,{f^{\prime }}\,{a_1}\,f
+ 15\,{f^{\prime }}\,{a_{0}}\,{f^{\prime \prime \prime }}^{2} +
180\,{f^{\prime }}\,{f^{\prime \prime }}^{2}\,{a_1}  +
180\,{f^{\prime \prime \prime }}\,{f^{\prime }}^{2}\,{a_1} - 3\,{a_{0}
}\,{f^{(v)}}\,{f^{\prime }}^{2} + 360\,{f^{\prime }}^{3}\,a_2
 + 15\,{f^{\prime }}\,{a_{0}}\,{{f^{(iv)}}}\,{f^{\prime \prime }}
-60\,
{f^{\prime \prime }}\,{a_1}\,{f^{\prime \prime \prime }}\,f
 - 12\,{a_{0}}\,{{f^{(iv)}}}\,{f^{\prime \prime \prime }}\,f + 240\,
{f^{\prime \prime \prime }}\,a_2\,f^{2} + {a_{0}}\,{f^{(v)}}\,{f^{\prime \prime }}\,f=0$. \hfill (A1.2)



These two equations determine $f$ in the sense that the complete set of
solutions for (\ref{eqgen}) is contained in the complete set of solutions which
satisfy both (A1.1) and (A1.2). The common set of solution to (A1.1) and
(A1.2) satisfy the following equation


$ {f^{\prime }}^{3}\,(3\,{f^{\prime \prime \prime }}\,{f^{\prime \prime }}^{2} - {f^{\prime \prime \prime }}^{
2}\,{f^{\prime }} - 3\,{f^{\prime }}\,{f^{\prime \prime }}\,{{f^{(iv)}}} +
{f^{(v)}}\,{f^{\prime }}^{2})\,( - {f^{\prime }}\,{f^{\prime \prime }} + f\,
{f^{\prime \prime \prime }})(2000\,f^{3}\,{f^{\prime }}^{5}\,{f^{\prime \prime }}^{6} + 1250
\,f^{7}\,{f^{\prime }}^{2}\,{f^{\prime \prime \prime }}^{5}
 - 18500\,f^{2}\,{f^{\prime }}^{7}\,{f^{\prime \prime }}^{5} + 50\,f
^{9}\,{f^{\prime \prime \prime }}^{4}\,{f^{(v)}} - 250\,{f^{\prime \prime \prime }}^{5}\,
{f^{\prime \prime }}\,f^{8} + 63000\,f\,{f^{\prime \prime }}^{4}\,{f^{\prime }}^{9}
 - 3000\,f^{7}\,{f^{\prime \prime \prime }}^{3}\,{f^{\prime \prime }}^{4}
 + 18\,f^{9}\,{f^{(v)}}^{3}\,{f^{\prime }}^{2} + 70875\,f
^{3}\,{f^{\prime }}^{8}\,{f^{\prime \prime \prime }}^{3} + 22500\,f^{5}\,{f^{\prime }}
^{5}\,{f^{\prime \prime \prime }}^{4} + 1692\,f^{6}\,{f^{\prime }}^{5}\,{{f^{(iv)}}}
^{3}
 + 150\,f^{9}\,{f^{\prime \prime \prime }}\,{f^{\prime }}\,{{f^{(iv)}}}^{2}\,
{f^{(v)}} - 70875\,{f^{\prime \prime }}^{3}\,{f^{\prime }}^{11} + 930\,f^{
8}\,{f^{\prime }}^{2}\,{f^{(v)}}\,{{f^{(iv)}}}^{2}\,{f^{\prime \prime }} -
5220\,f^{5}\,{f^{\prime }}^{6}\,{f^{(v)}}\,{f^{\prime \prime }}\,
{{f^{(iv)}}}
 +
2475\,f^{6}\,{f^{\prime \prime \prime }}^{2}\,{f^{\prime }}^{4}\,{f^{(v)}}\,{f^{\prime \prime }}
-
3300\,f^{6}\,{f^{\prime \prime \prime }}^{2}\,{f^{\prime }}^{3}\,{f^{\prime \prime }}^{2}\,{{f^{(iv)}}}
+32250\,f^{3}\,{f^{\prime }}^{7}\,{{f^{(iv)}}}\,{f^{\prime \prime }}^{3}
 - 10125\,f^{3}\,{f^{\prime }}^{8}\,{f^{\prime \prime }}^{2}\,{f^{(v)}} -
27000\,f^{2}\,{f^{\prime }}^{9}\,{f^{\prime \prime }}^{2}\,{{f^{(iv)}}} -
2875\,f^{6}\,{f^{\prime }}^{3}\,{f^{\prime \prime }}\,{f^{\prime \prime \prime }}^{4} -
9875\,f^{5}\,{f^{\prime }}^{3}\,{f^{\prime \prime }}^{4}\,{f^{\prime \prime \prime }}^{2}
 - 174375\,f^{2}\,{f^{\prime }}^{8}\,{f^{\prime \prime }}^{3}\,{f^{\prime \prime \prime }} +
212625\,f\,{f^{\prime }}^{10}\,{f^{\prime \prime }}^{2}\,{f^{\prime \prime \prime }} +
1875\,f^{4}\,{f^{\prime }}^{4}\,{f^{\prime \prime }}^{5}\,{f^{\prime \prime \prime }}
 + 182250\,f^{3}\,{f^{\prime }}^{7}\,{f^{\prime \prime }}^{2}\,{f^{\prime \prime \prime }}^{2} +
31000\,f^{3}\,{f^{\prime }}^{6}\,{f^{\prime \prime }}^{4}\,{f^{\prime \prime \prime }} -
1500\,f^{5}\,{f^{\prime \prime }}^{6}\,{f^{\prime }}^{2}\,{f^{\prime \prime \prime }} -
93375\,f^{4}\,{f^{\prime }}^{6}\,{f^{\prime \prime }}\,{f^{\prime \prime \prime }}^{3}
 - 8000\,f^{4}\,{f^{\prime }}^{5}\,{f^{\prime \prime }}^{3}\,{f^{\prime \prime \prime }}^{2} +
4750\,f^{6}\,{f^{\prime }}^{2}\,{f^{\prime \prime }}^{3}\,{f^{\prime \prime \prime }}^{3} +
5625\,f^{6}\,{f^{\prime }}\,{f^{\prime \prime }}^{5}\,{f^{\prime \prime \prime }}^{2} -
2875\,f^{5}\,{f^{\prime }}^{4}\,{f^{\prime \prime }}^{2}\,{f^{\prime \prime \prime }}^{3}
 + 375\,f^{7}\,{f^{\prime }}\,{f^{\prime \prime }}^{2}\,{f^{\prime \prime \prime }}^{4} -
212625\,f^{2}\,{f^{\prime }}^{9}\,{f^{\prime \prime }}\,{f^{\prime \prime \prime }}^{2} -
60\,f^{9}\,{f^{\prime \prime \prime }}^{2}\,{f^{\prime }}\,{f^{(v)}}^{2} +
90\,f^{9}\,{f^{(v)}}^{2}\,{f^{\prime \prime }}^{2}\,{f^{\prime \prime \prime }}
 - 300\,f^{8}\,{f^{\prime }}\,{{f^{(iv)}}}\,{f^{\prime \prime \prime }}^{4} +
750\,f^{8}\,{f^{\prime \prime \prime }}^{3}\,{f^{\prime \prime }}^{2}\,{{f^{(iv)}}} +
150\,f^{8}\,{f^{(v)}}\,{f^{\prime \prime }}^{3}\,{f^{\prime \prime \prime }}^{2} -
900\,f^{8}\,{f^{\prime \prime \prime }}\,{f^{\prime }}^{2}\,{{f^{(iv)}}}^{3}
 - 144\,f^{8}\,{f^{\prime }}^{3}\,{{f^{(iv)}}}\,{f^{(v)}}^{2} -
750\,f^{7}\,{f^{\prime \prime \prime }}^{3}\,{f^{\prime }}^{3}\,{f^{(v)}} +
5190\,f^{7}\,{f^{\prime \prime \prime }}^{2}\,{f^{\prime }}^{3}\,{{f^{(iv)}}}^{2} +
90\,f^{7}\,{f^{\prime \prime \prime }}\,{f^{\prime }}^{4}\,{f^{(v)}}^{2}
 - 1215\,f^{7}\,{f^{\prime }}^{3}\,{{f^{(iv)}}}^{3}\,{f^{\prime \prime }} -
441\,f^{7}\,{f^{\prime }}^{4}\,{{f^{(iv)}}}^{2}\,{f^{(v)}} +
30\,f^{7}\,{f^{\prime }}^{3}\,{f^{(v)}}^{2}\,{f^{\prime \prime }}^{2} -
14250\,f^{6}\,{f^{\prime \prime \prime }}^{3}\,{f^{\prime }}^{4}\,{{f^{(iv)}}}
 - 90\,f^{6}\,{f^{\prime }}^{5}\,{f^{\prime \prime }}\,{f^{(v)}}^{2} +
1875\,f^{6}\,{f^{\prime }}^{3}\,{f^{\prime \prime }}^{3}\,{{f^{(iv)}}}^{2} -
10125\,f^{5}\,{f^{\prime \prime \prime }}^{2}\,{f^{\prime }}^{6}\,{f^{(v)}} -
4185\,f^{5}\,{f^{\prime \prime \prime }}\,{f^{\prime }}^{6}\,{{f^{(iv)}}}^{2}
 - 1300\,f^{5}\,{f^{\prime }}^{4}\,{f^{\prime \prime }}^{4}\,{f^{(v)}} +
1500\,f^{5}\,{f^{\prime }}^{3}\,{{f^{(iv)}}}\,{f^{\prime \prime }}^{5} +
2355\,f^{5}\,{f^{\prime }}^{5}\,{{f^{(iv)}}}^{2}\,{f^{\prime \prime }}^{2} -
27000\,f^{4}\,{f^{\prime \prime \prime }}^{2}\,{f^{\prime }}^{7}\,{{f^{(iv)}}}
 + 4185\,f^{4}\,{f^{\prime }}^{7}\,{{f^{(iv)}}}^{2}\,{f^{\prime \prime }} -
9075\,f^{4}\,{f^{\prime }}^{5}\,{f^{\prime \prime }}^{4}\,{{f^{(iv)}}} +
6375\,f^{4}\,{f^{\prime }}^{6}\,{f^{(v)}}\,{f^{\prime \prime }}^{3} -
1815\,f^{7}\,{f^{\prime \prime \prime }}\,{f^{\prime }}^{3}\,{f^{\prime \prime }}\,{{f^{(iv)}}}\,{f^{(v)}}
 +
3240\,f^{7}\,{f^{\prime \prime \prime }}\,{f^{\prime }}^{2}\,{f^{\prime \prime }}^{2}\,{{f^{(iv)}}}^{2}
+750\,f^{7}\,{f^{\prime \prime \prime }}\,{f^{\prime \prime }}^{4}\,{f^{\prime }}\,{f^{(v)}} -
750\,f^{7}\,{f^{\prime }}^{2}\,{{f^{(iv)}}}\,{f^{\prime \prime }}^{3}\,{f^{(v)}}
 + 5220\,f^{6}\,{f^{\prime \prime \prime }}\,{f^{\prime }}^{5}\,{f^{(v)}}\,{{f^{(iv)}}} +
300\,f^{7}\,{f^{\prime \prime \prime }}^{2}\,{f^{\prime }}^{2}\,{f^{\prime \prime }}^{2}\,{f^{(v)}}
-
7500\,f^{6}\,{f^{\prime \prime \prime }}\,{f^{\prime }}^{2}\,{f^{\prime \prime }}^{4}\,{{f^{(iv)}}}
 +
25\,f^{6}\,{f^{\prime \prime \prime }}\,{f^{\prime }}^{3}\,{f^{(v)}}\,{f^{\prime \prime }}^{3}-
12180\,f^{6}\,{f^{\prime \prime \prime }}\,{f^{\prime }}^{4}\,{{f^{(iv)}}}^{2}\,{f^{\prime \prime }}
+
2865\,f^{6}\,{f^{\prime }}^{4}\,{f^{(v)}}\,{f^{\prime \prime }}^{2}\,{{f^{(iv)}}}
 +
18600\,f^{5}\,{f^{\prime \prime \prime }}\,{{f^{(iv)}}}\,{f^{\prime \prime }}^{3}\,{f^{\prime }}^{4}
-
8100\,f^{5}\,{f^{\prime \prime \prime }}\,{f^{(v)}}\,{f^{\prime }}^{5}\,{f^{\prime \prime }}^{2}
+120\,f^{8}\,{f^{\prime \prime \prime }}^{2}\,{f^{\prime }}^{2}\,{f^{(v)}}\,{{f^{(iv)}}}
 +
150\,f^{8}\,{f^{\prime \prime \prime }}^{2}\,{f^{\prime }}\,{{f^{(iv)}}}^{2}\,{f^{\prime \prime }}-
1080\,f^{8}\,{f^{\prime \prime \prime }}\,{f^{(v)}}\,{f^{\prime \prime }}^{2}\,{f^{\prime }}\,{{f^{(iv)}}}
+ 120\,f^{8}\,{f^{\prime \prime \prime }}\,{f^{(v)}}^{2}\,{f^{\prime }}^{2}\,{f^{\prime \prime }}
 - 150\,f^{9}\,{f^{\prime \prime \prime }}^{2}\,{f^{\prime \prime }}\,{f^{(v)}}\,{{f^{(iv)}}} -
90\,f^{9}\,{f^{\prime }}\,{f^{(v)}}^{2}\,{f^{\prime \prime }}\,{{f^{(iv)}}} +
54000\,f^{3}\,{f^{\prime \prime \prime }}\,{f^{\prime }}^{8}\,{f^{\prime \prime }}\,{{f^{(iv)}}} +
20250\,f^{4}\,{f^{\prime \prime \prime }}\,{f^{\prime }}^{7}\,{f^{(v)}}\,{f^{\prime \prime }}
 +
63900\,f^{5}\,{f^{\prime \prime \prime }}^{2}\,{f^{\prime }}^{5}\,{{f^{(iv)}}}\,{f^{\prime \prime }}
+250\,f^{8}\,{f^{\prime \prime \prime }}^{3}\,{f^{(v)}}\,{f^{\prime }}\,{f^{\prime \prime }} -
81900\,f^{4}\,{f^{\prime }}^{6}\,{{f^{(iv)}}}\,{f^{\prime \prime }}^{2}\,{f^{\prime \prime \prime }}
 - 1875\,f^{7}\,{f^{\prime \prime \prime }}^{3}\,{f^{\prime }}^{2}\,
{f^{\prime \prime }}\,{{f^{(iv)}}} +
975\,f^{7}\,{f^{\prime \prime \prime }}^{2}\,{f^{\prime }}\,{f^{\prime \prime }}^{3}\,{{f^{(iv)}}})=0.
\hfill $(A1.3)


Equation (A.13) has several factors. The only factor which provides
solutions to (\ref{eqgen}), when$n=3$ which are not of the form (1.5b) is


$(2000\,f^{3}\,{f^{\prime }}^{5}\,{f^{\prime \prime }}^{6} + 1250\,f^{7}\,
{f^{\prime }}^{2}\,{f^{\prime \prime \prime }}^{5} - 18500\,f^{2}\,{f^{\prime }}^{7}
\,{f^{\prime \prime }}^{5} - 250\,{f^{\prime \prime \prime }}^{5}\,{f^{\prime \prime }}\,f^{8} +
63000\,f\,{f^{\prime \prime }}^{4}\,{f^{\prime }}^{9}
 - 3000\,f^{7}\,{f^{\prime \prime \prime }}^{3}\,{f^{\prime \prime }}^{4} + 70875\,
f^{3}\,{f^{\prime }}^{8}\,{f^{\prime \prime \prime }}^{3} + 22500\,f^{5}\,
{f^{\prime }}^{5}\,{f^{\prime \prime \prime }}^{4} + 1692\,f^{6}\,{f^{\prime }}^{5}\,
{{f^{(iv)}}}^{3}
 - 70875\,{f^{\prime \prime }}^{3}\,{f^{\prime }}^{11} - 3300\,f^{6}
\,{f^{\prime \prime \prime }}^{2}\,{f^{\prime }}^{3}\,{f^{\prime \prime }}^{2}\,{{f^{(iv)}}}
 + 32250\,f^{3}\,{f^{\prime }}^{7}\,{{f^{(iv)}}}\,{f^{\prime \prime }}^{3} -
27000\,f^{2}\,{f^{\prime }}^{9}\,{f^{\prime \prime }}^{2}\,{{f^{(iv)}}}
 - 2875\,f^{6}\,{f^{\prime }}^{3}\,{f^{\prime \prime }}\,{f^{\prime \prime \prime }}
^{4} - 9875\,f^{5}\,{f^{\prime }}^{3}\,{f^{\prime \prime }}^{4}\,{f^{\prime \prime \prime }}
^{2} - 174375\,f^{2}\,{f^{\prime }}^{8}\,{f^{\prime \prime }}^{3}\,
{f^{\prime \prime \prime }} + 212625\,f\,{f^{\prime }}^{10}\,{f^{\prime \prime }}^{2}\,{f^{\prime \prime \prime }}
 + 1875\,f^{4}\,{f^{\prime }}^{4}\,{f^{\prime \prime }}^{5}\,{f^{\prime \prime \prime }} +
182250\,f^{3}\,{f^{\prime }}^{7}\,{f^{\prime \prime }}^{2}\,{f^{\prime \prime \prime }}^{2} +
31000\,f^{3}\,{f^{\prime }}^{6}\,{f^{\prime \prime }}^{4}\,{f^{\prime \prime \prime }} -
1500\,f^{5}\,{f^{\prime \prime }}^{6}\,{f^{\prime }}^{2}\,{f^{\prime \prime \prime }}
 - 93375\,f^{4}\,{f^{\prime }}^{6}\,{f^{\prime \prime }}\,{f^{\prime \prime \prime }}^{3} -
8000\,f^{4}\,{f^{\prime }}^{5}\,{f^{\prime \prime }}^{3}\,{f^{\prime \prime \prime }}^{2} +
4750\,f^{6}\,{f^{\prime }}^{2}\,{f^{\prime \prime }}^{3}\,{f^{\prime \prime \prime }}^{3} +
5625\,f^{6}\,{f^{\prime }}\,{f^{\prime \prime }}^{5}\,{f^{\prime \prime \prime }}^{2}
 - 2875\,f^{5}\,{f^{\prime }}^{4}\,{f^{\prime \prime }}^{2}\,{f^{\prime \prime \prime }}^{3} +
375\,f^{7}\,{f^{\prime }}\,{f^{\prime \prime }}^{2}\,{f^{\prime \prime \prime }}^{4} -
212625\,f^{2}\,{f^{\prime }}^{9}\,{f^{\prime \prime }}\,{f^{\prime \prime \prime }}^{2} -
300\,f^{8}\,{f^{\prime }}\,{{f^{(iv)}}}\,{f^{\prime \prime \prime }}^{4}
 + 750\,f^{8}\,{f^{\prime \prime \prime }}^{3}\,{f^{\prime \prime }}^{2}\,{{f^{(iv)}}} -
900\,f^{8}\,{f^{\prime \prime \prime }}\,{f^{\prime }}^{2}\,{{f^{(iv)}}}^{3} +
5190\,f^{7}\,{f^{\prime \prime \prime }}^{2}\,{f^{\prime }}^{3}\,{{f^{(iv)}}}^{2} -
1215\,f^{7}\,{f^{\prime }}^{3}\,{{f^{(iv)}}}^{3}\,{f^{\prime \prime }}
 - 14250\,f^{6}\,{f^{\prime \prime \prime }}^{3}\,{f^{\prime }}^{4}\,{{f^{(iv)}}} +
1875\,f^{6}\,{f^{\prime }}^{3}\,{f^{\prime \prime }}^{3}\,{{f^{(iv)}}}^{2} -
4185\,f^{5}\,{f^{\prime \prime \prime }}\,{f^{\prime }}^{6}\,{{f^{(iv)}}}^{2} +
1500\,f^{5}\,{f^{\prime }}^{3}\,{{f^{(iv)}}}\,{f^{\prime \prime }}^{5}
 + 2355\,f^{5}\,{f^{\prime }}^{5}\,{{f^{(iv)}}}^{2}\,{f^{\prime \prime }}^{2} -
27000\,f^{4}\,{f^{\prime \prime \prime }}^{2}\,{f^{\prime }}^{7}\,{{f^{(iv)}}} +
4185\,f^{4}\,{f^{\prime }}^{7}\,{{f^{(iv)}}}^{2}\,{f^{\prime \prime }} -
9075\,f^{4}\,{f^{\prime }}^{5}\,{f^{\prime \prime }}^{4}\,{{f^{(iv)}}}
 +
3240\,f^{7}\,{f^{\prime \prime \prime }}\,{f^{\prime }}^{2}\,{f^{\prime \prime }}^{2}\,{{f^{(iv)}}}^{2}
-
7500\,f^{6}\,{f^{\prime \prime \prime }}\,{f^{\prime }}^{2}\,{f^{\prime \prime }}^{4}\,{{f^{(iv)}}}
-
12180\,f^{6}\,{f^{\prime \prime \prime }}\,{f^{\prime }}^{4}\,{{f^{(iv)}}}^{2}\,{f^{\prime \prime }}
 +
18600\,f^{5}\,{f^{\prime \prime \prime }}\,{{f^{(iv)}}}\,{f^{\prime \prime }}^{3}\,{f^{\prime }}^{4}
+150\,f^{8}\,{f^{\prime \prime \prime }}^{2}\,{f^{\prime }}\,{{f^{(iv)}}}^{2}\,{f^{\prime \prime }}
+ 54000\,f^{3}\,{f^{\prime \prime \prime }}\,{f^{\prime }}^{8}\,{f^{\prime \prime }}\,{{f^{(iv)}}}
 +
63900\,f^{5}\,{f^{\prime \prime \prime }}^{2}\,{f^{\prime }}^{5}\,{{f^{(iv)}}}\,{f^{\prime \prime }}
-
81900\,f^{4}\,{f^{\prime }}^{6}\,{{f^{(iv)}}}\,{f^{\prime \prime }}^{2}\,{f^{\prime \prime \prime }}
-
1875\,f^{7}\,{f^{\prime \prime \prime }}^{3}\,{f^{\prime }}^{2}\,{f^{\prime \prime }}\,{{f^{(iv)}}}
 + 975\,f^{7}\,{f^{\prime \prime \prime }}^{2}\,{f^{\prime }}\,{f^{\prime \prime }}^{
3}\,{{f^{(iv)}}}) + ( - 750\,f^{7}\,{f^{\prime \prime \prime }}^{3}\,{f^{\prime }}^{ 3} -
1300\,f^{5}\,{f^{\prime }}^{4}\,{f^{\prime \prime }}^{4} - 10125\,f^{3}
\,{f^{\prime }}^{8}\,{f^{\prime \prime }}^{2}
 + 150\,f^{8}\,{f^{\prime \prime }}^{3}\,{f^{\prime \prime \prime }}^{2} - 10125\,f
^{5}\,{f^{\prime \prime \prime }}^{2}\,{f^{\prime }}^{6} - 441\,f^{7}\,{f^{\prime }}
^{4}\,{{f^{(iv)}}}^{2} + 6375\,f^{4}\,{f^{\prime }}^{6}\,{f^{\prime \prime }} ^{3} +
50\,f^{9}\,{f^{\prime \prime \prime }}^{4}
 + 150\,f^{9}\,{f^{\prime \prime \prime }}\,{f^{\prime }}\,{{f^{(iv)}}}^{2}
 + 930\,f^{8}\,{f^{\prime }}^{2}\,{{f^{(iv)}}}^{2}\,{f^{\prime \prime }} -
5220\,f^{5}\,{f^{\prime }}^{6}\,{f^{\prime \prime }}\,{{f^{(iv)}}} + 2475\,f
^{6}\,{f^{\prime \prime \prime }}^{2}\,{f^{\prime }}^{4}\,{f^{\prime \prime }}
 + 750\,f^{7}\,{f^{\prime \prime \prime }}\,{f^{\prime \prime }}^{4}\,{f^{\prime }}
 - 750\,f^{7}\,{f^{\prime }}^{2}\,{{f^{(iv)}}}\,{f^{\prime \prime }}^{3} +
5220\,f^{6}\,{f^{\prime \prime \prime }}\,{f^{\prime }}^{5}\,{{f^{(iv)}}} + 300\,f^{
7}\,{f^{\prime \prime \prime }}^{2}\,{f^{\prime }}^{2}\,{f^{\prime \prime }}^{2}
 + 25\,f^{6}\,{f^{\prime \prime \prime }}\,{f^{\prime }}^{3}\,{f^{\prime \prime }}^{3
} + 2865\,f^{6}\,{f^{\prime }}^{4}\,{f^{\prime \prime }}^{2}\,{{f^{(iv)}}} -
8100\,f^{5}\,{f^{\prime \prime \prime }}\,{f^{\prime }}^{5}\,{f^{\prime \prime }}^{2} + 120
\,f^{8}\,{f^{\prime \prime \prime }}^{2}\,{f^{\prime }}^{2}\,{{f^{(iv)}}}
 - 150\,f^{9}\,{f^{\prime \prime \prime }}^{2}\,{f^{\prime \prime }}\,{{f^{(iv)}}}
 + 20250\,f^{4}\,{f^{\prime \prime \prime }}\,{f^{\prime }}^{7}\,{f^{\prime \prime }} + 250
\,f^{8}\,{f^{\prime \prime \prime }}^{3}\,{f^{\prime }}\,{f^{\prime \prime }} - 1815\,f^{7}
\,{f^{\prime \prime \prime }}\,{f^{\prime }}^{3}\,{f^{\prime \prime }}\,{{f^{(iv)}}}
 - 1080\,f^{8}\,{f^{\prime \prime \prime }}\,{f^{\prime \prime }}^{2}\,{f^{\prime }}
\,{{f^{(iv)}}}){f^{(v)}} + ( - 90\,f^{9}\,{f^{\prime }}\,
{f^{\prime \prime }}\,{{f^{(iv)}}} + 30\,f^{7}\,{f^{\prime }}^{3}\,{f^{\prime \prime }}^{2} -
144\,f^{8}\,{f^{\prime }}^{3}\,{{f^{(iv)}}}
 - 90\,f^{6}\,{f^{\prime }}^{5}\,{f^{\prime \prime }} +
90\,f^{9}\,{f^{\prime \prime }}^{2}\,{f^{\prime \prime \prime }} -
60\,f^{9}\,{f^{\prime \prime \prime }}^{2}\,{f^{\prime }} +
120\,f^{8}\,{f^{\prime \prime \prime }}\,{f^{\prime }}^{2}\,{f^{\prime \prime }} +
90\,f^{7}\,{f^{\prime \prime \prime }}\,{f^{\prime }}^{4}){f^{(v)}}^{2} + \\
18\,f^{9}\,{f^{\prime }}^{2}\,{f^{(v)}}^{3}=0 $. \hfill (A1.4)
}
\pagebreak

\renewcommand{\theequation}{B1.\arabic{equation}}
\section{}

In this appendix we show that $g\left( x_{1},x_{2},x_{3}\right) $ given by
(\ref{5d2}) acts as a distribution. For convenience we write $x_{1}=x,\;x_{2}=y$
and $x_{3}=0,$ so that we consider
\begin{equation}
g\left( x,y\right) =g_{1}\left( x,y\right) +g_{2}\left( x,y\right)
+g_{3}(x,y),
\end{equation}
where
\begin{equation}
g_{1}\left( x,y\right) =\dfrac{-2}{\left( x-y\right) ^{2}x^{3}}-\dfrac{2}{
\left( x-y\right) ^{3}x^{2}},
\end{equation}
\begin{equation}
g_{2}\left( x,y\right) =\dfrac{-2}{y^{3}\left( x-y\right) ^{2}}+\dfrac{2}{
y^{2}\left( x-y\right) ^{3}},
\end{equation}
and
\begin{equation}
g_{3}\left( x,y\right) =\dfrac{2}{x^{3}y^{2}}+\dfrac{2}{x^{2}y^{3}}.
\end{equation}
We wish to prove that
\begin{equation}
g\left( x,y\right) =\pi ^{2}\left( \delta ^{\prime \prime }\left( x\right)
\delta ^{\prime }\left( y\right) +\delta ^{\prime \prime }\left( y\right)
\delta ^{\prime }\left( x\right) \right).
\end{equation}
To do this we first regularise integrals which have singularities, using the
Hademard finite part interpretation of these integrals. We will see that
this is consistent with the usual definition of generalised Fourier
transforms used in \S 4 and also with the definition of $\widehat{f}\left(
k\right) $ in terms of $\widehat{f}_{U}\left( k\right) $ and
$\widehat{f}_{L}\left( k\right) ,$ see (\ref{6d10}).

For a function $\phi \left( x\right) ,$ which is bounded in the neighbourhood of
the origin, we define
\begin{equation}
^{^{^{\displaystyle{\ast}}}}\!\!\!\int_{-\infty }^{\infty }\dfrac{\phi \left( x\right) }{x^{2}}dx=\underset{
\epsilon \rightarrow 0}{\lim }\left( \int\limits_{\left| x\right| \geq
\epsilon }\dfrac{
\phi \left( x\right) }{x^{2}}dx-\dfrac{2\phi \left( 0\right) }{\epsilon }\right)
\end{equation}
and
\begin{equation}
^{^{^{\displaystyle{\ast}}}}\!\!\!\int_{-\infty }^{\infty }\dfrac{\phi \left( x\right) }{x^{3}}dx=\underset{
\epsilon \rightarrow 0}{\lim }\left( \int\limits_{\left| x\right| \geq
\epsilon }\dfrac{
\phi \left( x\right) }{x^{3}}dx-\dfrac{2\phi ^{\prime }\left( 0\right)
}{\epsilon }\right) .
\end{equation}
When $\phi \left( x\right) =e^{-ikx}$ we have, from the above definitions
\begin{equation}
^{^{^{\displaystyle{\ast}}}}\!\!\!\int_{-\infty }^{\infty }\dfrac{e^{-ikx}}{x^{2}}dx=-\pi \left|
k\right| ,
\end{equation}
corresponding to the interpretation of the Fourier transform of $1/x^{2}$ in
\S 3 and \S 5.

Now we let $\phi \left( x,y\right) $ be a test function and define
\begin{equation}
^{^{^{\displaystyle{\ast}}}}\!\!\!\int_{-\infty }^{\infty }\dfrac{\phi \left( x,y\right) }{x^{2}\left(
x-y\right) ^{3}}dy=\underset{\epsilon \rightarrow 0}{\lim }\left(
\int\limits_{\left| x-y\right| \geq \epsilon }\dfrac{\phi \left(
x,y\right) }{x^{2}\left( x-y\right) ^{3}}dy+\dfrac{2}{\epsilon }\dfrac{
\phi _{2}\left( x,x\right) }{x^{2}}\right),
\end{equation}
where $\phi _{2}\left( x,y\right)=\dfrac{\partial }{\partial y}$, and hence
\begin{multline}
^{^{^{\displaystyle{\ast \ast}}}}\!\!\!\int_{-\infty }^{\infty }\int_{-\infty }^{\infty }\dfrac{\phi
\left( x,y\right) }{x^{2}\left( x-y\right) ^{3}}dxdy=\underset{\epsilon
\rightarrow 0}{\lim }  \Bigg( \;\;\;\; \iint\limits_{\left| x\right| \geq
\epsilon \left|, x-y\right| \geq \epsilon} \dfrac{\phi\left( x,y\right) }{x^{2}
\left( x-y\right) ^{3}}dy \\
+\dfrac{2}{\epsilon } \int\limits_{\left| y\right| \geq \epsilon }\dfrac{\phi 
\left( 0,y\right) }{y^{3}}dy+\dfrac{2}{\epsilon }\int\limits_{\left| x\right|
\geq \epsilon }\dfrac{\phi _{2}\left( x,x\right) }{x^{2}}dx
-\dfrac{4}{ \epsilon ^{2}}\phi _{2}\left( 0,0\right) \Bigg) .
\end{multline}
Similarly we define
\begin{equation}
^{^{^{\displaystyle{\ast}}}}\!\!\!\int_{-\infty }^{\infty }\dfrac{\phi \left( x,y\right) }{x^{3}\left(
x-y\right) ^{2}}=\underset{\epsilon \rightarrow 0}{\lim }\left(
\int\limits_{\left| x-y\right| \geq \epsilon }\dfrac{\phi \left(
x,y\right) }{x^{3}\left( x-y\right) ^{2}}dy-\dfrac{2}{\epsilon }\dfrac{
\phi \left( x,x\right) }{x^{3}}\right) .
\end{equation}
Now since
\begin{equation}
\int\limits_{\left| x,y\right| \geq \epsilon }\dfrac{\phi \left(
x,y\right) dy}{x^{3}\left( x-y\right) ^{2}}=\int\limits_{\left| \eta
\right| \geq \epsilon }\dfrac{\phi \left( x,\eta +x \right) d\eta
}{x^{3}\eta ^{2}}\equiv \dfrac{1}{x^{3}}I,
\end{equation}
with
\begin{equation}
\dfrac{\partial I}{\partial x}=\int\limits_{\left| \eta \right| \geq
\epsilon }\dfrac{\left( \phi _{1}\left( x,\eta +x\right) +\phi
_{2}\left( x,\eta +x \right) \right) d\eta }{\eta ^{2}},
\end{equation}
where again $\phi _{1}\left( x,y\right)=\dfrac{\partial }{\partial x}
 \phi \left(x,y\right)$,  we can, with the use of (B1.17) define
\begin{multline}
^{^{^{\displaystyle{\ast \ast}}}}\!\!\!\int_{-\infty }^{\infty }\int_{-\infty }^{\infty }\dfrac{\phi
\left( x,y\right) }{x^{3}\left( x-y\right) ^{2}}dxdy=\underset{\epsilon
\rightarrow 0}{\lim }\iint\limits_{\left| x\right| \geq \epsilon
\left| x-y\right|, \geq \epsilon}   \dfrac{\phi \left( x,y\right) }{
x^{3}\left( x-y\right) ^{3}}dxdy  \\
-\dfrac{2}{\epsilon }\int\limits_{\left| \eta \right| \geq \epsilon }
\dfrac{\phi _{1}\left( 0,\eta \right) +\phi _{2}\left( 0,\eta \right) dy}{
\eta ^{2}}-\dfrac{2}{\epsilon }\int\limits_{\left| x\right| \geq
\epsilon }\dfrac{\phi \left( x,x\right) dx}{x^{3}}+\dfrac{4}{\epsilon }
\left( \phi _{1}\left( 0,0\right) +\phi _{2}\left( 0,0\right) \right) .
\end{multline}

Using similar definitions for the remaining four integrals arising in $\iint
\phi g$ we may define
\begin{multline}
^{^{^{\displaystyle{\ast \ast}}}}\!\!\!\iint\limits_{\mathbb{R}^{2}}\phi gdxdy
=\underset{\epsilon \rightarrow 0}{\lim }\Bigg(
\iint\limits_{\Omega _{1}}\phi g_{1}dxdy+\iint\limits_{\Omega
_{2}}\phi g_{2}dxdy+\iint\limits_{\Omega _{3}}\phi g_{3}dxdy \\
+
\dfrac{4}{\epsilon ^{2}}( \phi ( \epsilon ,\epsilon )
-\phi ( -\epsilon ,-\epsilon ) -\phi ( \epsilon ,0) +\phi (-\epsilon ,0) 
-\phi (0,\epsilon )+\phi ( 0,-\epsilon ) )\Bigg)
\end{multline}
This requires a good deal of simplification in integration by parts. The
regions $\Omega _{1},\Omega _{2}$ and $\Omega _{3}$ appearing in (B1.15) are
given by
\begin{equation}
\begin{array}{clllcrc}
\Omega _{1}= & \{\left| x\right| \geq \epsilon  & , & \left| x-y\right| \geq
\epsilon  & , &
\left| y\right| <\epsilon \} & , \\
\Omega _{2}= & \{\left| x-y\right| \geq \epsilon  & , & \left| y\right| \geq
\epsilon  & , &
\left| x\right| <\epsilon \} & , \\
\Omega _{3}= & \{\left| x\right| \geq \epsilon  & , & \left| y\right| \geq
\epsilon  & , &
\left| x-y\right| <\epsilon \} & .
\end{array}
\end{equation}
This simplifies to
\begin{equation}
^{^{^{\displaystyle{\ast \ast}}}}\!\!\!\iint\limits_{R^{2}}\phi gdxdy=\underset{\epsilon \rightarrow 0}{\lim }
\left( \sum\limits_{j=1}^{3}\iint\limits_{\Omega _{j}}\phi g_{j}dxdy\right)
+4\phi _{xxy}\left( 0,0\right) +4\phi _{yyx}\left( 0,0\right) .
\end{equation}
A lengthy, but straightforward calculation shows that the limit of the sum
is
$$\left( \pi ^{2}-4\right) \left( \phi _{xxy}\left( 0,0\right) +\phi _{yyx}
\left(0,0\right) \right) .$$
Hence
\begin{equation}
^{^{^{\displaystyle{\ast \ast}}}}\!\!\!\iint\limits_{R^{2}}\phi g=\pi ^{2}\left( \phi _{xxy}\left( 0,0\right)
+\phi _{yyx}\left( 0,0\right) \right) .
\end{equation}
This confirms that the function $g\left( x,y\right) $ acts as the
distribution given by (B1.5).

\end{document}